\documentclass[twocolumn,superscriptaddress,floatfix,
showpacs,10pt]{revtex4}
\usepackage[english]{babel}
\usepackage{graphicx}
\usepackage{amsmath}
\usepackage{amssymb}
\usepackage{bm}
\usepackage{color}

\begin{document}

\title{Confinement-deconfinement transition due to spontaneous symmetry breaking in quantum Hall bilayers}

\author{D. I. Pikulin}
\affiliation{Department of Physics and Astronomy, University of
British Columbia, Vancouver, BC, Canada V6T 1Z1}
\affiliation{Quantum Matter Institute, University of British Columbia, Vancouver BC, Canada V6T 1Z4}
\affiliation{Instituut-Lorentz, Universiteit Leiden, P.O. Box 9506, 2300 RA Leiden, The Netherlands}
\author{P. G. Silvestrov}
\affiliation{Institute for Mathematical Physics, TU Braunschweig, 38106 Braunschweig, Germany}
\author{T. Hyart}
\affiliation{Department of Physics and Nanoscience Center, University of Jyv\"askyl\"a, P.O. Box 35 (YFL), FI-40014 University of Jyv\"askyl\"a, Finland}
\affiliation{Instituut-Lorentz, Universiteit Leiden, P.O. Box 9506, 2300 RA Leiden, The Netherlands}

\begin{abstract}
Band-inverted electron-hole bilayers support quantum spin Hall insulator and exciton condensate phases. We investigate such a bilayer in an external magnetic field. We show that the interlayer correlations lead to formation of a  {\it helical quantum Hall exciton condensate state}. In contrast to the chiral edge states of the quantum Hall exciton condensate in electron-electron bilayers, existence of the  counterpropagating edge modes results in formation of a ground state spin-texture not supporting gapless single-particle excitations. This feature has deep consequences for the low energy behavior of the system. Namely, the charged edge excitations in a sufficiently narrow Hall bar are {\it confined} i.e.~a charge on one of the edges always gives rise to an opposite charge on the other edge. Moreover, we show that magnetic field and gate voltages allow to control confinement-deconfinement transition of charged edge excitations, which can be probed with nonlocal conductance. Confinement-deconfinement transitions are of great interest, not least because of their possible significance in shedding light on the confinement problem of quarks.
\end{abstract}

\maketitle

The role of the electron-electron interactions for the
experimentally accessible topological media is best appreciated in
quantum Hall (QH) systems. The fractionally charged quasiparticles
have been  studied  at the
fractional filling factors \cite{DasSarma08, Jain07}, and the
non-abelian excitations of more exotic QH states may eventually
lead to a revolution in quantum computing \cite{Read00, Nayak08,
Lindner12, Clarke13, Mong14}. However, Coulomb interactions play a
crucial role also in the case of the integer filling factors
\cite{Chakraborty87, Fertig89, Sondhi93, Moon95, DasSarma08,
Eisenstein04, Girvin99}. Remarkably, interactions create a QH
ferromagnetic ground state at $\nu=1$ even in the absence of
Zeeman energy. In such systems, the $SU(2)$ spin rotation symmetry
is spontaneously broken, resulting in the low-energy excitations
being spin waves and charged topological spin textures, skyrmions
\cite{Moon95, DasSarma08, Girvin99, Sondhi93}. The presence of a
small symmetry-breaking Zeeman field does not change the
low-energy excitations qualitatively.

In QH bilayer systems the role of spin is played by the layer
index (pseudospin)  \cite{Chakraborty87, Fertig89, Moon95,
Girvin99}. In this case the $SU(2)$ pseudospin rotation symmetry
is explicitly broken by the interactions, as they are
larger within the layers than between the layers. The interactions
favor the pseudospin orientations in $(x,y)$-plane, where the direction is chosen spontaneously
(spontaneous $U(1)$ symmetry breaking) so that the QH bilayers
realize an easy-plane ferromagnet. Since the spontaneously
chosen direction in the $(x,y)$-plane corresponds to a spontaneous
interlayer phase coherence, this easy-plane ferromagnetic state is
equivalent to an exciton condensate  \cite{Fertig89, Moon95}.

\begin{figure}[t]
\includegraphics[width = \linewidth]{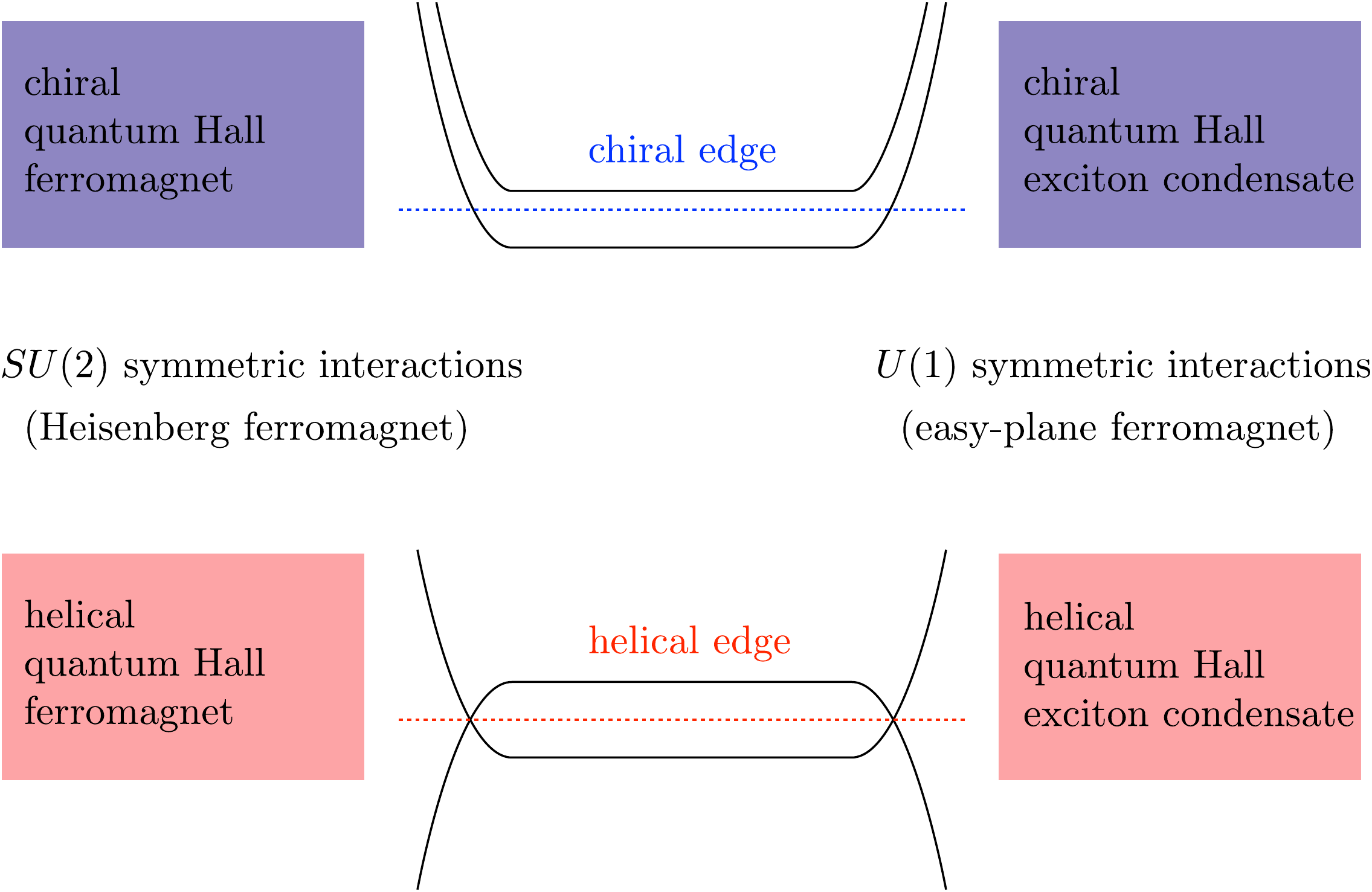}
\caption{Four different types of experimentally accessible QH pseudospin ferromagnetic states at $\nu=1$. The single layer realizations (left column) realize Heisenberg ferromagnets because the interactions have $SU(2)$ symmetry. The bilayer QH exciton condensates (right column) can be described as an easy-plane ferromagnet with a spontaneously broken $U(1)$-symmetry. We argue that the classification is additionally enriched as the QH systems can support either chiral (top row) or helical (bottom row) edge excitations.
}\label{fig:bilayers}
\end{figure}

The QH ferromagnet and QH exciton condensate in electron-electron
bilayers support a single chiral edge mode. However the
two Landau levels may also support counterpropagating edge modes.
The natural hosts of such kind of QH states are systems supporting quantum spin Hall (QSH) effect
\cite{Kane05, Bernevig06, Konig07, Liu08, Du13, Spanton14} due to
inverted electron-hole bandstructure. In
these materials the magnetic field allows to tune through the
Landau level crossing \cite{Scharf12, Pikulin14}, where we
expect to find a QH state with spontaneously-broken
(pseudo)spin-rotation symmetry. Thus, we argue that there exist
four different experimentally accessible pseudospin ferromagnetic
states, determined by spontaneously broken symmetry [SU(2) in
single layer and U(1) in bilayer systems] and the edge structure
[chiral or helical]. All these possibilities are illustrated in
 Fig.~\ref{fig:bilayers}.

In this paper we concentrate on the helical QH exciton condensate
state [broken U(1) symmetry and helical edge structure].
Remarkably, we find that in this system the charged edge
excitations in a sufficiently narrow Hall bar are confined: A
charge on one of the edges is always connected to the opposite
charge on the other edge through the bulk by a stripe of rotated
pseudospins, and thus low-energy isolated charged excitations
cannot be observed. The gapless single-particle 
excitations are prohibited 
since the electron-electron interactions lead to an edge reconstruction and opening
of a single-particle gap \cite{Iordanskii99,Fertig06}. However,
unlike it happens in the existing
examples, the helical exciton condensate creates long-range
correlations between edges. We show that a magnetic
field and gate voltages can be used to tune in and out of the
exciton condensate phase. Thus this system provides a unique
opportunity to study a confinement-deconfinement transition,
similar to the one which is hypothesized to liberate the quarks
from their color confinement at extremely high temperatures or
densities \cite{Greensite11}. Finally, we show that the confined
and deconfined phases can be distinguished using
nonlocal conductance.

\section{Helical quantum Hall exciton condensate phase}

\begin{figure}
\includegraphics[width = \linewidth]{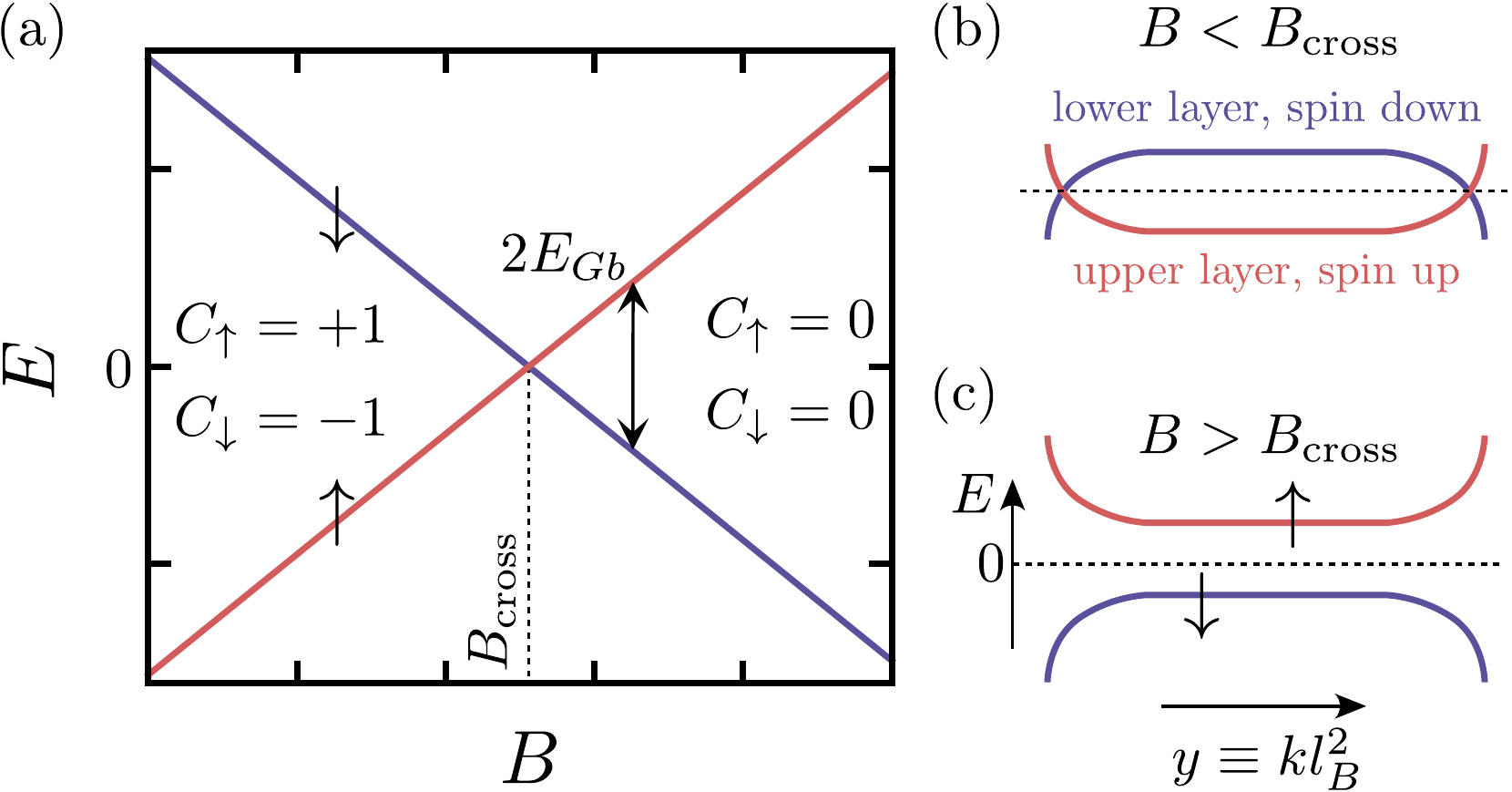}
\caption{Magnetic field and momentum dependence of the lowest Landau level energies. (a) There exists a robust crossing of the lowest Landau level energies as a function of magnetic field at $B=B_{\rm cross}$, because the energy of the electron-like Landau level with spin up (red line) increases and the energy of the hole Landau level with spin down (blue line) decreases as a function of $B$. We denote the energy separation between these two levels as $2 E_{Gb}$. (b-c) Momentum (or equivalently position) dependencies of the lowest Landau level energies. The electron-like (hole-like) Landau level bends upwards (downwards) in energy near the edge of the sample. (b) For $B<B_{\rm cross}$ the system supports helical edge modes protected by spin-resolved Chern numbers  $C_\uparrow=-C_\downarrow=1$. (c)  For $B>B_{\rm cross}$ the edge is gapped according to the non-interacting theory ($C_\uparrow=C_\downarrow=0$).
}\label{fig:landau_levels_en_mom}
\end{figure}

We consider bilayer QSH systems, such as InAs/GaSb \cite{Liu08, Du13, Spanton14}, described by the BHZ Hamiltonian \cite{Bernevig06, Liu08, supplementary}. The important property of these systems is that  there is a crossing of electron and hole Landau levels as a function of magnetic field at $B=B_{\rm cross}$ \cite{Scharf12, Pikulin14, supplementary} [see Fig.~\ref{fig:landau_levels_en_mom}(a)], where the band inversion is removed. Near this crossing the  single-particle Hamiltonian is  \cite{supplementary}
 \begin{equation}
\hat{H}_0=\sum_{k} [\hat{\psi}_{k, \uparrow}^\dag \hat{\psi}_{k, \uparrow} - \hat{\psi}_{k, \downarrow}^\dag \hat{\psi}_{k, \downarrow} ] E_G(kl_B^2),
\label{LLham}
\end{equation}
where $E_G(y) = E_G(k l_B^2)$ is the energy-momentum dispersion of Landau levels and $\hat{\psi}^\dag_{k\uparrow (\downarrow)}$ are the electron creation operators for the lowest electron and hole Landau levels. Here we have fixed the total filling factor of the Landau levels $\nu_T=\nu_\uparrow+\nu_\downarrow=1$, and utilized the fact that the momentum $k$ in the Landau level wavefunctions is directly connected to the position $y$ in the real space. Importantly, the spin and layer degrees of freedom are locked with each other, so that the pseudospin $\uparrow (\downarrow)$ means simultaneously up (down) spin and upper (lower) layer \cite{supplementary}.  The Fermi level is set to be at zero energy.

In the bulk the energy $E_G(y)=E_{Gb}$ is independent of the momentum ($E_{Gb}<0$ for $B < B_{\rm cross}$ and $E_{Gb} > 0$ for $B > B_{\rm cross}$). When approaching the edge the Landau level originating from the electron (hole) band always disperses upwards (downwards) in energy. The spatial variation of $E_G(y)$ occurs within a characteristic length scale $l_0\gtrsim l_B$, which depends on the details of the edge, but due to topological reasons $E_G(y)>0$ reaches extremely large values (on the order of the energy separation between the bulk Landau levels) close to the edge \cite{supplementary}. Therefore, for the magnetic fields $B < B_{\rm cross}$, $E_G(y)$ goes through zero near the edge, yielding to the helical edge states [see Fig.~\ref{fig:landau_levels_en_mom}(b)]. On the other hand for $B>B_{\rm cross}$ the edge is gapped in the non-interacting theory [see Fig.~\ref{fig:landau_levels_en_mom}(c)].

The electron-electron interactions $\hat{H}_I$ are described by
\begin{equation}
\hat{H}_I= \frac{1}{2} \sum_{\sigma, \sigma'} \sum_{k, k', q} V_P^{\sigma \sigma'}(k-k', q) \hat{\psi}_{k\sigma}^\dag \hat{\psi}_{k', \sigma'}^\dag \hat{\psi}_{k'+q \sigma'} \hat{\psi}_{k-q \sigma},
\end{equation}
where $V_P^{\sigma \sigma'}(k-k', q)$ is obtained by projecting the Coulomb interactions to the subspace generated by the wavefunctions of the lowest Landau levels \cite{supplementary}. Here, we assumed that the higher Landau levels are energetically separated from the lowest ones by an energy gap larger than the characteristic energy scale of the Coulomb interactions $V_C=e^2/(4 \pi \epsilon \epsilon_0 l_B)$. We find that this assumption can be satisfied with the material parameters corresponding to InAs/GaSb bilayers \cite{Liu13}.

To find the ground state of the Hamiltonian
$\hat{H}=\hat{H}_0+\hat{H}_I$, we consider states where the local
direction of the pseudospin $\mathbf{h}(\mathbf{r})$
($|\mathbf{h}(\mathbf{r})| = 1$) varies in space. Because the Hamiltonian
is translationally invariant in the $x$-direction, we assume that $\mathbf{h}(\mathbf{r})$ is independent of $x$ \cite{comment}. By further noticing that
the $y$-dependence translates to a momentum dependence of the
pseudospin $h_i(y)=h_i(kl_B^2)$, we can express our ansatz for the
ground state many-particle wavefunction as a Slater determinant
$|\Psi[\mathbf{h}(kl_B^2)]\rangle$, where for each momentum $k$ we
create an electron with pseudospin pointing along
$\mathbf{h}(kl_B^2)$. To compute the ground state, we need to
minimize the energy functional for such kind of pseudospin texture
\cite{Moon95}. For the energy functional we obtain
\cite{supplementary}
 \begin{eqnarray}
E &=& E_0 -  \sum_{k, k'}  \bigg\{ \sum_{i=x,y} V_P^{XY}(k-k') h_i(k l_B^2) h_i(k' l_B^2)   \nonumber \\
   && \hspace{-0.5cm} + V_P^Z(k-k')  h_z(k l_B^2)h_z(k' l_B^2) \bigg\}  + \sum_k E_G(kl_B^2)h_z(k l_B^2). \nonumber\\ \label{energyfunctional}
 \end{eqnarray}
Here $V^Z_P(q)=[-V_P^{\uparrow\uparrow}(q, 0)+V_P^{\uparrow\uparrow}(q, q)+V_P^{\uparrow \downarrow}(q, 0)]/4$ and $V_P^{XY}(q)=V_P^{\uparrow\downarrow}(q, q)/4$ are the interaction coefficients, which characterize the anisotropy of the interactions within a layer and between the layers.

\begin{figure}[t]
\includegraphics[width = 0.93 \linewidth]{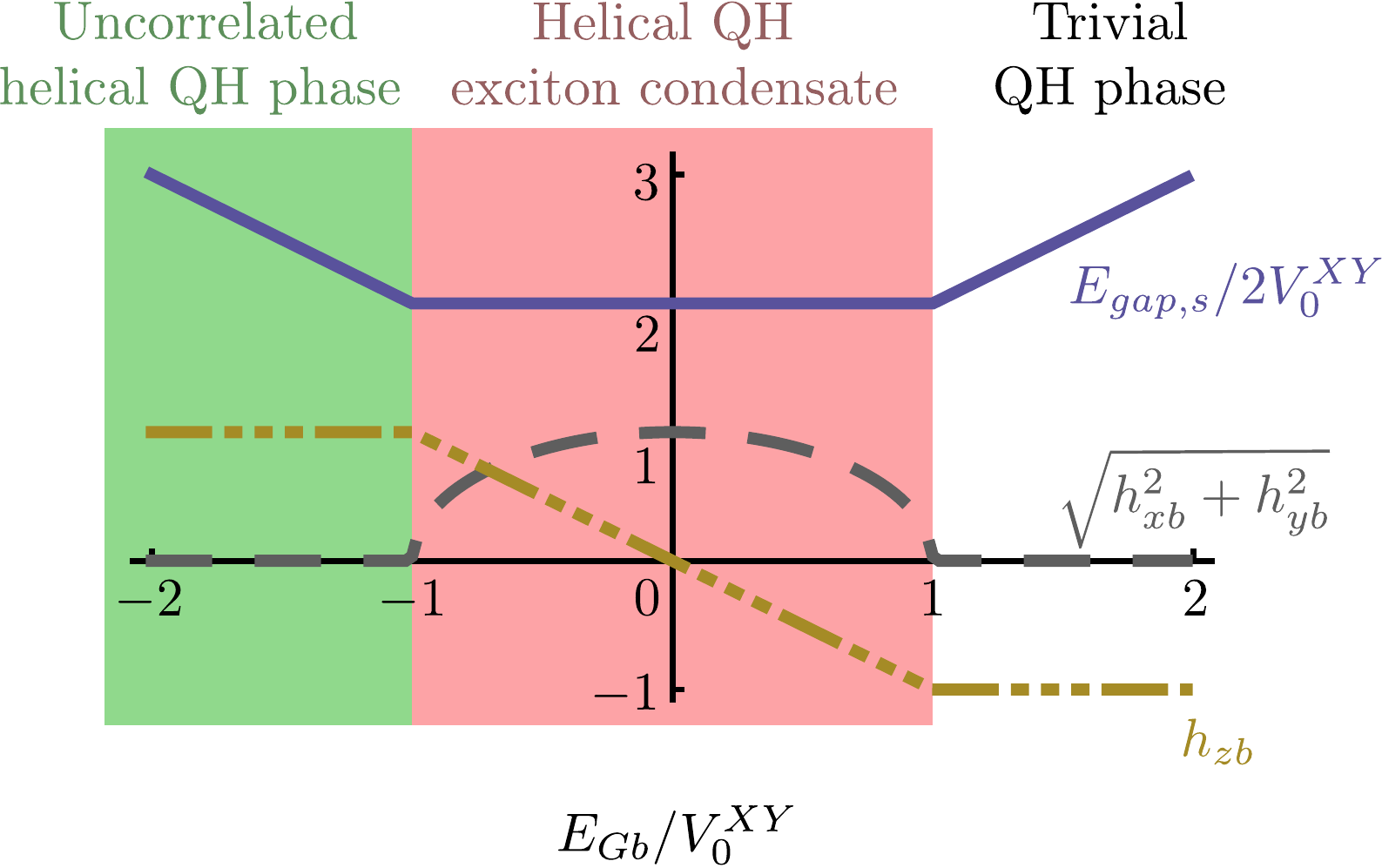}
\caption{Phase diagram. For $E_{Gb}<-2(V^{XY}_0-V^Z_0)$ the system supports \textit{an uncorrelated helical QH phase}. In this phase, there is no spontaneous interlayer phase coherence in the bulk ($h_{xb}^2+h_{yb}^2=0$) and the system supports helical edge states (guaranteed by spin-resolved Chern numbers $C_{\uparrow/\downarrow}=\pm1$).  For $|E_{Gb}|<2(V^{XY}_0-V^Z_0)$ the system supports \textit{a helical QH exciton condensate phase}. In this phase, there is spontaneous interlayer phase coherence ($h_{xb}^2+h_{yb}^2 \ne 0$), and the system supports exotic confined edge excitations (see below). For $E_{Gb}>2(V^{XY}_0-V^Z_0)$ the system is in \textit{a trivial QH phase}, where the edge is fully gapped. $E_{Gb}$ can be controlled with gate voltages or magnetic field \cite{comment2}. Charged bulk excitations are gapped  everywhere in the phase diagram ($E_{\rm gap, s}>0$). We have chosen $V_0^{Z}/V_0^{XY}=1/2$. \label{fig:phasediagram}}
\end{figure}

We start by considering an infinite system. In this case, the pseudospin direction  $\mathbf{h}(\mathbf{r})$ is spatially homogeneous. By minimizing the energy-functional (\ref{energyfunctional}), we find  that the pseudospin direction  $\mathbf{h}_b$ is determined by the parameters $E_{Gb}$ and $V^{Z (XY)}_0 = \sum_q V^{Z (XY)}_P (q)$  \cite{supplementary}. Here
$E_{Gb}$ acts as an effective magnetic field preferring the pseudospin direction along $- {\rm sgn}(E_{Gb}) \hat{e}_z$.  On the other hand, the interactions prefer the pseudospin directions within the (x,y)-plane ($V^{XY}_0>V^Z_0$), and the energy cost to rotate the pseudospin so that it points along the $z$-direction is proportional to $V^{XY}_0-V^Z_0$. Thus, as a balance between these two competing effects, the direction of the pseudospin is tilted away from the $(x,y)$-plane, resulting in the three distinct phases of the
system, which are summarized in Fig.~\ref{fig:phasediagram}. For
sufficiently large $|E_{Gb}|$, we see that $|h_{z b}|=1$, meaning
that only one layer is occupied.
The phases $h_{zb}=1$
({\it uncorrelated helical QH phase}) and $h_{zb}=-1$ ({\it
trivial QH phase}) are topologically distinct from each other. For
$E_{Gb}<-2(V^{XY}_0-V^Z_0)$ and $h_{zb}=1$ the system supports
helical edge modes (the spin-resolved Chern numbers are
$C_{\uparrow/\downarrow}=\pm1$). On the other hand, in the regime
$E_{Gb}>2(V^{XY}_0-V^Z_0)$ and $h_{zb}=-1$, the edge is completely
gapped (the spin-resolved Chern numbers are
$C_{\uparrow/\downarrow}= 0$). Between these two phases is the
helical QH exciton condensate phase, where $|h_{z b}| < 1$ and
thus $h_{xb}^2 + h_{yb}^2 \neq 0$. In this phase the direction of
the pseudospin projection onto the $(x, y)$-plane is determined
spontaneously. Because the pseudospin in
this system labels spin and layer index simultaneously, this phase
has simultaneously spontaneous in-plane spin polarization and
spontaneous interlayer phase coherence.

\begin{figure*}[ht]
\includegraphics[width = \linewidth]{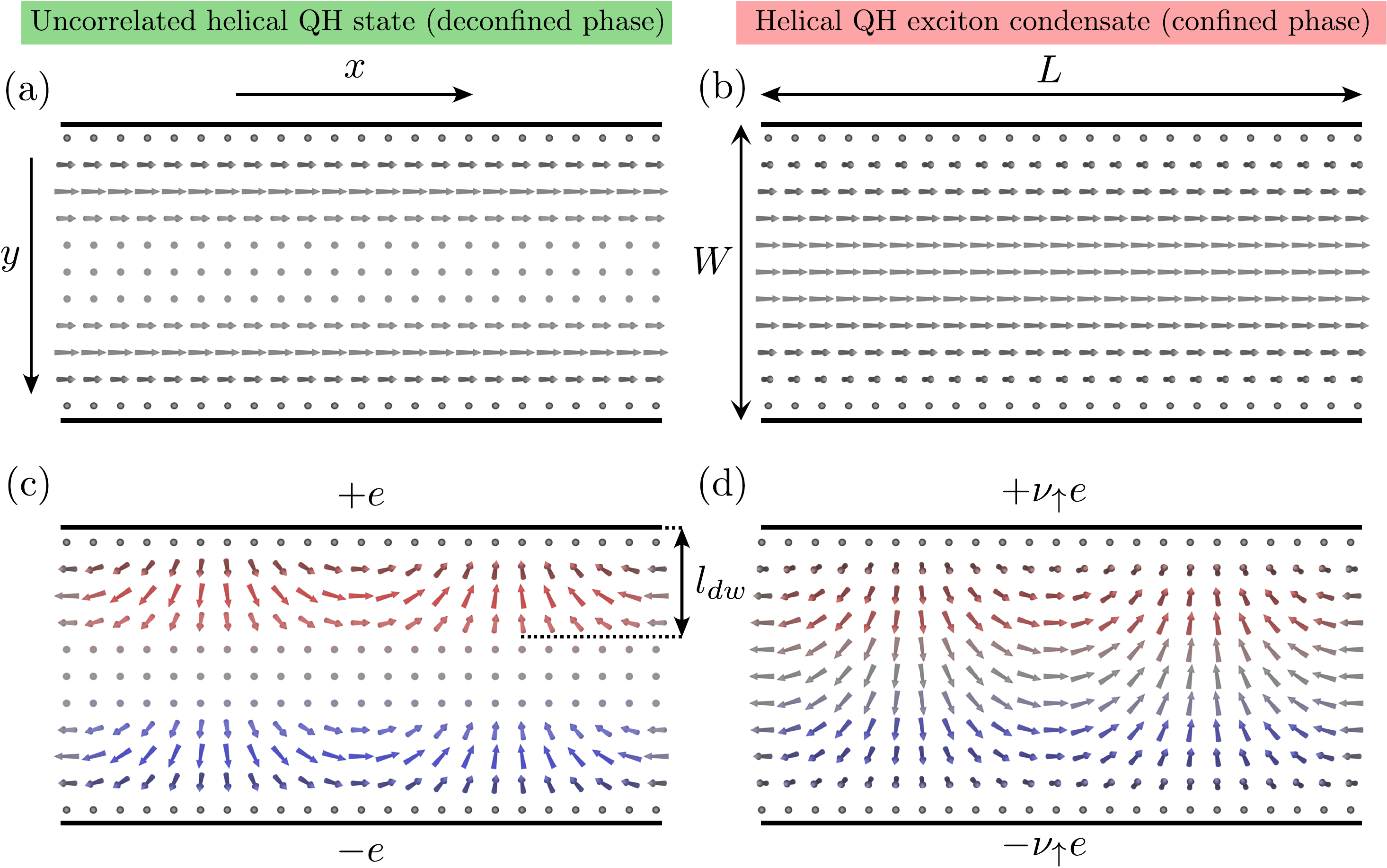}
\caption{Spin textures. (a),(b) Ground state spin textures for the uncorrelated helical QH phase and the helical QH exciton condensate phase, respectively. In both cases there exists a robust domain wall, where the polar angle of the pseudospin magnetization $\theta(y)$ rotates from $\pi$ to $\theta_b$ along the $y$-direction. In the uncorrelated phase $\theta_b=0$ but in the helical QH exciton condensate phase $\theta_b \ne 0, \pi$ indicating spontaneous interlayer phase coherence in the bulk. In both cases the ground states are degenerate for all choices of the constant azimuthal angle $\phi$ of the pseudospin magnetization. (We have chosen $\phi=0$.) (c),(d) Charged excitations can be created by letting the azimuthal angle $\phi(x)$ to rotate along the $x$-direction.  In a closed system (obtained by connecting the ends of the sample to form a narrow cylinder) $\phi(x)$ must rotate integer multiples of $2 \pi$. The energy to create such kind of excitation in the uncorrelated helical QH phase scales as $\delta E \sim V_2^{XY} l_{dw}/L$, and the elementary excitations, which carry charge $\pm e$, can be created independently on the different edges. On the other hand, deep inside the helical QH exciton condensate phase $\delta E \sim V_2^{XY} W/L$ and the elementary charges are $\pm \nu_{\uparrow}e$. Furthermore, the charged edge excitations are confined: a charge  $\pm \nu_{\uparrow}e$ on one of the edges is always connected to the opposite charge on the other edge by a stripe of rotated pseudospins through the bulk, and thus isolated charges cannot be observed at low energies. The  charge density obtained from Eq.~(\ref{eq:charge_density}) is shown with red (positive charge density) and blue (negative charge density) colors.}\label{fig:excitations}
\end{figure*}

The bulk gap for single particle excitations $E_{\rm gap, s}$ can be calculated using Hartree-Fock linearization \cite{supplementary}
\begin{equation}
E_{\rm gap, s}=2 \sqrt{(E_{Gb}-2 V^Z_0 h_{zb})^2+ 4 (V^{XY}_0)^2 (h_{xb}^2+h_{yb}^2)}. \nonumber
\end{equation}
Additionally to the single particle excitations, the helical QH exciton condensate supports collective excitations \cite{Moon95}: the neutral pseudospin waves (Goldstone mode)  give rise to spin and counterflow charge superfluidity, and the lowest energy charged excitations are topological pseudospin textures, which carry fractional charge $\pm \nu_{\uparrow (\downarrow)} e$. Here $\nu_{\uparrow/\downarrow}=(1\pm h_{zb})/2$ are the pseudospin resolved filling factors of the different Landau levels. The energy required to create these charged  excitations is slightly lower than $E_{\rm gap, s}$ \cite{Moon95}.

 We point out that although $h_{zb}=\pm 1$ are topologically distinct phases, the bulk gap for creating charged excitations never closes, when one tunes from one phase to the other by controlling $E_{Gb}$. This is possible because the pseudospin rotation symmetry, which protects the existence of spin-resolved Chern numbers as topological numbers, is spontaneously broken in the helical exciton condensate phase. The interacting BHZ model for bilayers shows somewhat similar behavior also at zero magnetic field, where a trivial insulator phase can be connected to a quantum spin Hall insulator phase without closing of the bulk gap, because of an intermediate phase where the time-reversal symmetry is spontaneously broken \cite{Pikulin14a}. It is also experimentally known that the exciton condensate phase with $\nu_{\uparrow}=\nu_{\downarrow}=1/2$ can be smoothly connected to uncorrelated QH state with  $\nu_{\uparrow}=1$ and $\nu_{\downarrow}=0$ in conventional QH bilayers \cite{Ding}. Experimental investigations of InAs/GaSb bilayers in the QH regime \cite{Du13, Nichele14} are consistent with this prediction, because no gap closing has been observed as a function of magnetic field.

\section{Confinement-deconfinement transition of edge excitations}

\begin{figure}[t]
\includegraphics[width = 0.93\linewidth]{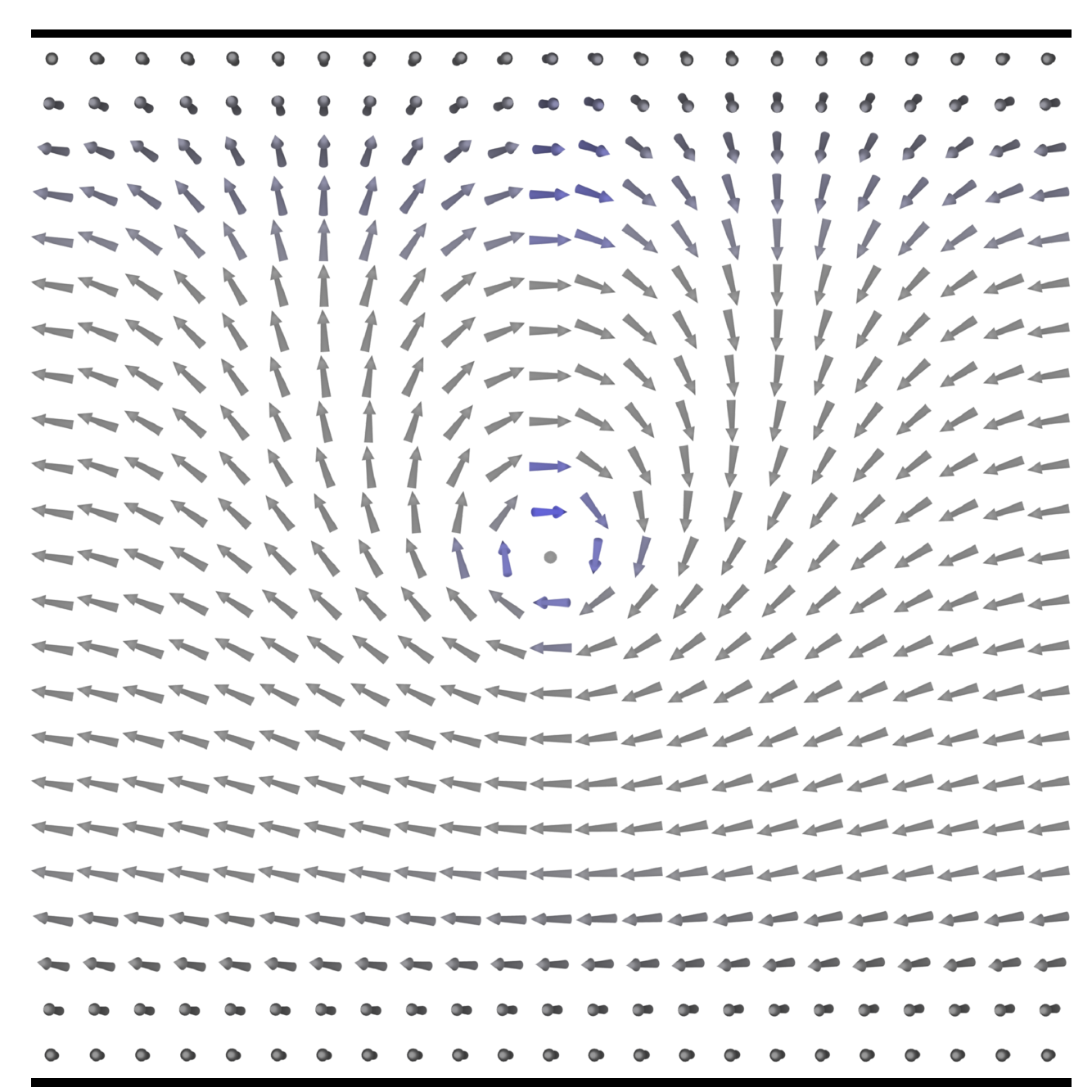}
\caption{Charged excitation localized at one edge of the sample in the helical QH exciton condensate. The excitation can be visualized as a meron-antimeron pair, where the meron is localized inside the sample and anti-meron -- outside. These excitations become the lowest energy charged excitations in sufficiently wide samples (otherwise similar cylinder geometry as in Fig.~\ref{fig:excitations}), where it is energetically favorable to break the stripe connecting opposite charges at the different edges by a creation of a bulk meron. Depending on the profile $E_{G}(y)$ near the edge, the lowest energy excitation is charge-neutral or having charge $\pm e$. The charge is determined by the pseudospin orientation in the center of the meron. The energy of such excitation is of order of Coulomb energy $V_0^{XY}$. }
\label{fig:bulkmeron}
\end{figure}

We now turn to the description of the ground state pseudospin
texture $h_z(y)=\cos [\theta_0(y)]$,
$h_{x}(y)=\sin [\theta_0(y)] \cos(\phi)$ and $h_{y}(y)=\sin
[\theta_0(y)] \sin(\phi)$ at the edge. (The ground state will be degenerate
with respect to the choice of $\phi$.) As discussed above close to
the edge $E_G(y)>0$ takes large values, because the edge states
are topologically protected to exist at all energies between the
lowest Landau levels and higher ones. Therefore close to the edge
$\theta_0(y)=\pi$. On the other hand, in the bulk
$\theta_0(y)=\theta_b=\arccos(h_{zb})$. This means that there
always exists a domain wall, where $\theta_0$ rotates from $\pi$
to $\theta_b$. Although the existence of the domain wall is a
robust topological property of the system, the detailed shape of
$\theta_0(y)$ and the length scale $l_{dw}$, where this rotation
happens, depend on the details of the sample \cite{supplementary}.
The ground states in the uncorrelated helical QH phase ($\theta_b=0$) and helical
QH exciton condensate phase ($\theta_b \ne 0, \pi$) are illustrated in
Fig.~\ref{fig:excitations} (a) and (b), respectively.  It turns out that the existence of spontaneous interlayer phase coherence, which distinguishes the two
different phases of matter, also has deep consequences on the
nature of the low-energy excitations in this system.

By using a Hartree-Fock linearization for the ground state, we find that the single particle excitations are gapped also close to the edge, and the magnitude of the energy gap is determined by the Coulomb energy scale $V_0^{XY}$. However, similarly as for the case of a coherent domain wall in QH ferromagnetic state in graphene \cite{Fertig06}, the lowest energy edge excitations are not the single-particle ones. Namely, the ground state is degenerate with respect to the choice of $\phi$, and therefore in accordance with the Goldstone's theorem the system supports low-energy excitations described by spatial variation of $\phi(\mathbf{r})$. Due to the general relationship between the electric and topological charge densities  in QH ferromagnets \cite{Moon95, supplementary, Fertig06}
\begin{equation}
\delta \rho(\mathbf{r})=-\frac{e}{4 \pi} \frac{\partial \phi}{\partial x} \frac{\partial \cos[\theta_0(y)]}{\partial y}\label{eq:charge_density}
\end{equation}
these excitations also carry charge, which is localized at the edges of the sample. In this section we illustrate these excitations in a closed system obtained by connecting the ends of the sample to form a narrow cylinder with width $W$ and circumference $L$  \cite{comment3}.

We start by considering this kind of closed system, where $W \ll L$ (see Fig.~\ref{fig:excitations}).  This geometry is topologically equivalent to a Corbino ring, which has been experimentally realized for QH exciton condensates \cite{Finck11, Xuting12}. Using Eqs.~(\ref{energyfunctional}) and (\ref{eq:charge_density}) we find   that the lowest energy excitations correspond to rotation of $\phi(x)$ by $2\pi$ and carry a net charge within one of the edges \cite{supplementary}. They have an energy
$\delta E \sim V_2^{XY} l_{dw}/L$ in the uncorrelated phase and $\delta E \sim V_2^{XY} W/L$ deep in the helical exciton condensate phase. Here $V^{XY}_2=\frac{1}{2} \sum_{q} V_P^{XY}(q) q^2 l_B^2$ characterizes the cost of exchange energy caused by  $\nabla \phi(x)$ \cite{supplementary}.

In the uncorrelated phase these excitations have a charge $\pm e$. By inspecting Fig.~\ref{fig:excitations}, we  notice that because the spin points along $z$-direction in the bulk, there is no rotation happening in the bulk. This means that we can choose separate fields $\phi_1(x)$ and  $\phi_2(x)$ for the two edges, so that these excitations can be created independently on the different edges much as in graphene \cite{Fertig06}.

The situation is dramatically different in the helical QH exciton
condensate phase. There,  the elementary excitations in a
closed system have a charge $\pm \nu_\uparrow e$. Moreover, as
illustrated in Fig.~\ref{fig:excitations}, a charge  $\pm
\nu_{\uparrow}e$ on one of the edges is always connected to the
opposite charge on the other edge by a stripe of rotated bulk
pseudospins. Breaking the bulk pseudospin configuration costs an
energy comparable to the Coulomb energy, and thus isolated charges
cannot be observed at low energies. This means that this type of
charged edge excitations in the helical QH exciton condensate
phase are confined.

It is illustrative to consider what happens to the excitations in
the helical QH exciton condensate,  when the width of the  sample
is increased. Namely, the excitation energy increases
proportionally to the width of the sample  $\delta E \sim V_2^{XY} W/L$ and eventually for $W
\sim L$ it becomes energetically favorable instead of having a
large area of rotating spins between two edges to create a bulk
meron (see Fig.~\ref{fig:bulkmeron}). This resembles the physics
of quarks, where the growing separation of a quark-antiquark pair
eventually results in the creation of a new quark-antiquark pair between them.

\section{Luttinger liquid theory and nonlocal transport}

\begin{figure}[t]
\includegraphics[width = 0.8\linewidth]{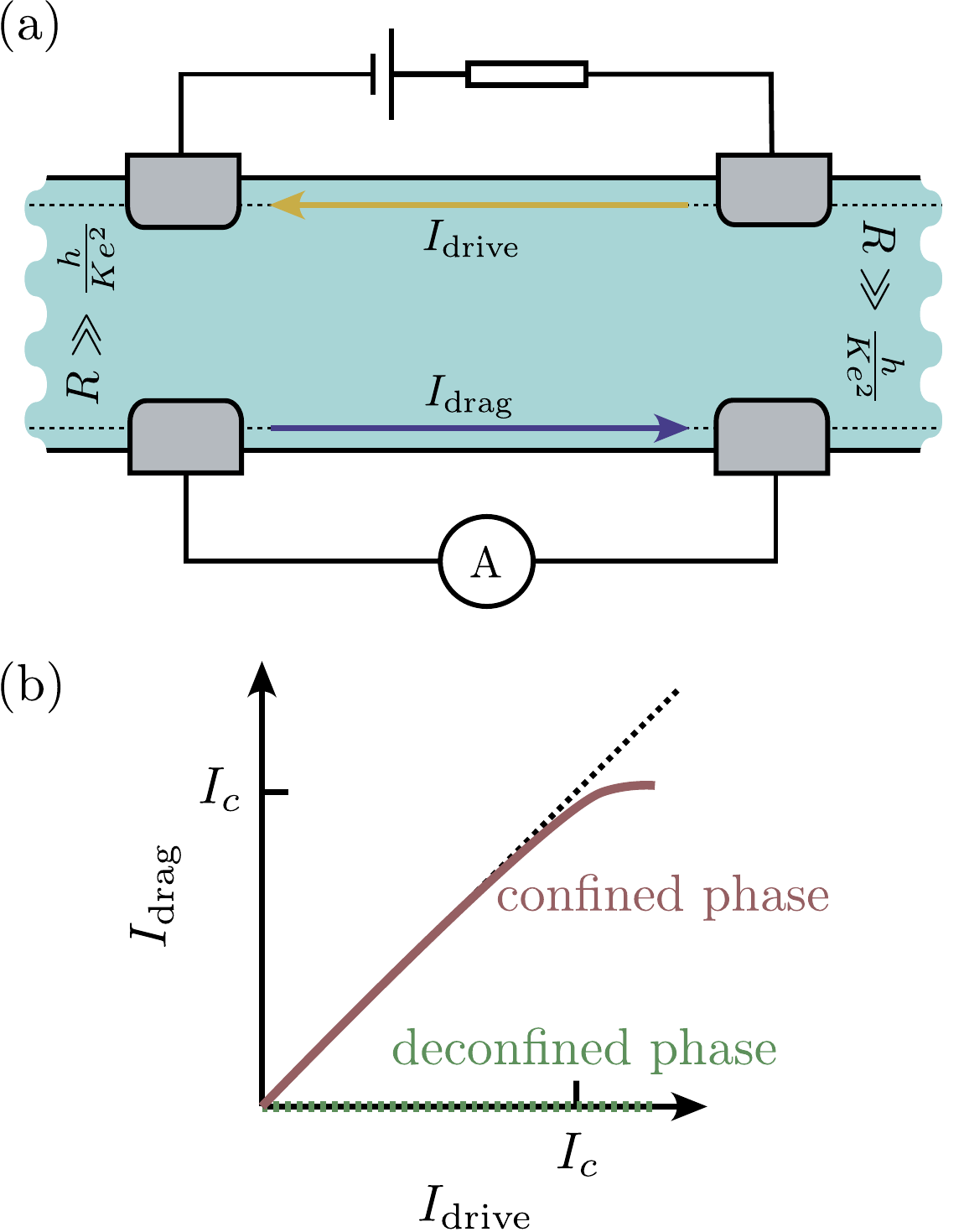}
\caption{Nonlocal transport to demonstrate the charge
confinement. (a) The transport geometry. A drive current is applied on one of the edges and
the resulting drag current is measured on the opposite edge.
(b) Due to the charge confinement in the helical QH exciton
condensate phase, the drive current necessarily gives rises to an
opposite drag current on the other edge. On the other hand, in the
uncorrelated (deconfined) phase, there is only a weak drag current
due to the Coulomb force acting between the charges.
}\label{fig:measurement}
\end{figure}

To predict experimentally measurable consequences of the charge
confinement, we consider nonlocal transport in an open system. By
considering the time-dependent field theory for the pseudospin for
a reasonably narrow sample in the helical QH exciton condensate
phase we arrive at an effective one-dimensional Hamiltonian
\cite{supplementary}
 \begin{equation}
H  = \int dx \bigg[ \frac{e^2}{2 W \hbar^2 \Gamma} \Pi(x)^2 +
\frac{W \rho_{sb}}{2} (\partial_x \phi(x))^2 \bigg],
 \end{equation}
where $\rho_{sb}=V^{XY}_2 \sin^2\theta_b/\pi $ is the pseudospin stiffness and $\Gamma=e^2/(16 \pi l_B^2 (V_0^{XY}-V_0^Z))$ describes the interlayer capacitance per unit area, which is strongly enhanced from the electrostatic value by the exchange interactions. The one-dimensional charge densities in the different edges (labeled 1 and 2)
$\rho_{1, 2}(x) = \mp \frac{e}{4\pi}(1+ \cos \theta_b)\frac{\partial \phi}{\partial x}$
are always opposite and determined by a single field $\phi(x)$, highlighting the confinement of the charged edge excitations. The one-dimensional theory describes a Luttinger liquid, and the so-called Luttinger parameter $K$ in the convention used in Ref.~\cite{Giamarchi}, is given by
 \begin{align}
K=\frac{l_B}{W} \sqrt{\frac{V_0^{XY} - V_0^Z}{V_2^{XY}}} \frac{(1+\cos\theta_b)^2}{4 \sin\theta_b}.
 \nonumber
 \end{align}
The Luttinger parameter in quantum Hall systems determines the
conductance for ideal contacts $G_{cf}=Ke^2/h$ \cite{Giamarchi,
KaneFisher}. Because the pseudospin waves are charge neutral in the
bulk, conductance decreases with $W$ as  $G_{cf} \propto
1/W$. It is important to notice that in this system $G_{cf}$
describes the conductance for a counterflow/drag geometry,  where
opposite currents are flowing in the two edges. The helical QH
exciton condensate phase does not support net transport current as
long as the voltages $eV$ are small compared to $\hbar v \pi/W$
\cite{supplementary}. This automatically leads to a remarkable
transport property that characterizes the helical QH exciton
condensate phase.  Namely, by considering a nonlocal transport
geometry shown in Fig.~(\ref{fig:measurement}) (a), where a drive
current is applied on one of the edges and a resulting drag
current is measured on the opposite edge, we find that necessarily
$I_{{\rm drag}}=I_{{\rm drive}}$ at small voltages. This should be
contrasted to the uncorrelated helical phase, where the charged
edge excitations are deconfined. In that case, one has two
independent Luttinger liquid theories for the two edges
\cite{supplementary}, and therefore one expects only a weak drag
current due to the Coulomb force acting between the charges. For
$W \gg l_B$, we expect that this effect is negligible compared to
drag current in the confined phase.

Finally, to estimate the critical current $I_c$, where the relation $I_{{\rm drag}}=I_{{\rm drive}}$ breaks down, we notice that the maximum voltage is determined by the gap $e V_{max} \approx \hbar v\pi/W$. By using reasonable estimates $V_0^Z = V_0^{XY}/2$, $V_2^{XY}=V_0^{XY}/4$, $\theta_b=\pi/2$, $W=20 l_B$, $l_B=10$ nm, $v=14$ km/s \cite{Spielman, Hyart},
we find $I_c = G_{cf} V_{max} \approx 0.1$ nA.

\section{Summary and discussion}

In summary, we have predicted the existence of a helical QH
exciton condensate state in band-inverted electron-hole bilayers.
We have shown that the counterpropagating edge modes give rise to a
ground state pseudospin texture, where the polar angle of the
pseudospin magnetization $\theta(y)$ rotates from the boundary
value $\pi$ to the bulk value $\theta_b$ along the direction
perpendicular to the edge. Low-energy charged
excitations can be created by letting the azimuthal angle of the
pseudospin polarization $\phi(x)$ to rotate along the edge.
Remarkably, in a sufficiently narrow Hall bar these charged edge
excitations are confined in the presence of spontaneous interlayer
phase-coherence ($\theta_b\ne 0, \pi$): a charge on one of the
edges always gives rise to the opposite charge on the other edge,
and thus isolated charges cannot be observed at low energies. 
Moreover, we
predict the possibility to control $\theta_b$  with a magnetic
field and gate voltages. This allows to study a
confinement-deconfinement transition, which occurs
simultaneously with the bulk phase-transition between the helical
QH exciton condensate phase ($\theta_b\ne 0, \pi$) and the
uncorrelated helical QH phase ($\theta_b=0$).

The helical QH exciton condensate phase can be experimentally probed using Josephson-like interlayer tunneling and counterflow superfluidity \cite{DasSarma08, Girvin99, Eisenstein04, Moon95, Finck11, Xuting12, Spielman, Hyart}. Moreover, because the pseudospin in this system describes simultaneously both the spin and the layer degrees of freedom, the helical QH exciton condensate phase can also be probed using the spin superfluidity and the NMR techniques  \cite{Girvin99}. Perhaps it is even possible to use local probe techniques to image the confinement-deconfinement transition and the confinement physics as illustrated in Figs.~\ref{fig:excitations} and \ref{fig:bulkmeron}. Finally, we have shown that the charge confinement also gives rise to a remarkable new transport property. Namely, a drive current applied on one of the edges gives rise to  exactly opposite drag current $I_{{\rm drag}}=I_{{\rm drive}}$ at the other edge.

Our results for the confinement of the edge excitations may also
be applicable to the so-called canted antiferromagnetic phase,
which is predicted to appear in graphene \cite{Murthy14}.
Similarly to the helical QH exciton condensate state considered in
this paper, the canted antiferromagnetic phase is
characterized by a spontaneously broken U(1)-symmetry in the bulk
and a single edge supports only gapped meron-antimeron excitations
\cite{Murthy14}.

The phenomena of confinement stemming from the particle physics
models \cite{Greensite11} has been studied  also in condensed
matter systems
\cite{Volovik-book, Volovik-review,Helium-confinement-experiment, Yang96,
Lake10}. However, we expect that the combination of the different techniques for probing the helical QH exciton condensate phase will provide a more
intuitive understanding and new perspectives on the confinement physics.

We also point out that InAs/GaSb bilayers is a promising system for superconducting applications, and edge-mode superconductivity has already been experimentally demonstrated in the QSH regime \cite{Pribiag}. In the presence of superconducting contacts, the helical QH exciton condensate may provide a new route for realizing exotic nonlocal Josephson effects and non-Abelian excitations, such as parafermions \cite{Lindner12, Clarke13}.

{\it Acknowledgements.--} This work was supported by the Academy of Finland Center of Excellence program, the European Research Council (Grant No. 240362-Heattronics), the Dutch Science Foundation NWO/FOM, NSERC, CIfAR, Max Planck - UBC Centre for Quantum Materials, and the DFG grant RE 2978/1-1.

\onecolumngrid

\newpage

\setcounter{section}{0}
\renewcommand*{\thesection}{}
\section*{Supplementary material for "Confinement-deconfinement transition due to spontaneous symmetry breaking in quantum Hall bilayers"}
\section{Effective model from the BHZ Hamiltonian}

\begin{figure}[h]
\includegraphics[width = 0.4\linewidth]{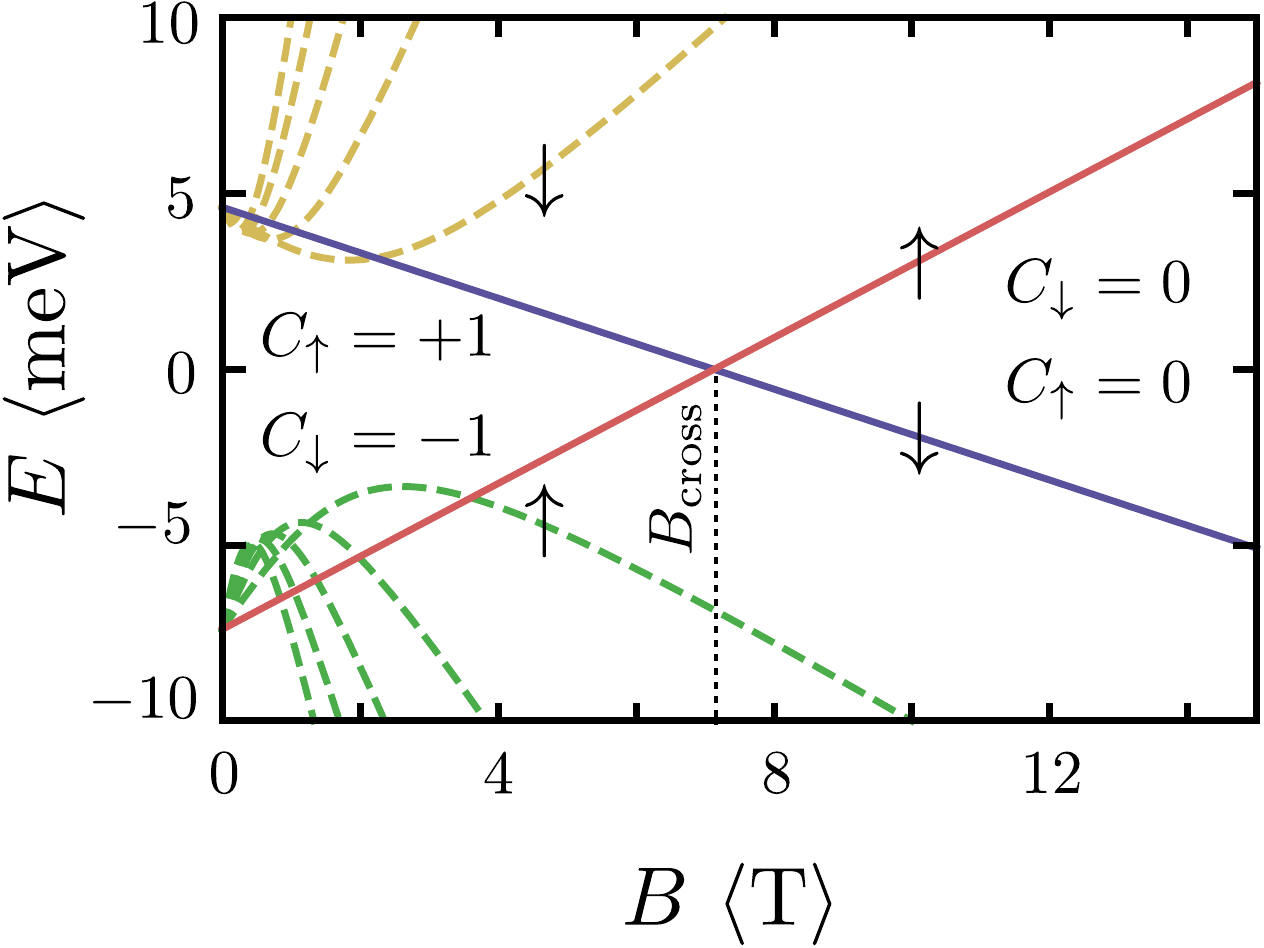}
\caption{Landau level fan for the InAs/GaSb bilayer. The figure shows the energies of the lowest 6 Landau levels on each side of the Fermi level (in terms of the minimum of the absolute value of energy as a function of magnetic field) for the Hamiltonian \eqref{eq:single_particle} with characteristic values of the parameters for InAs/GaSb bilayers ${\cal{M}}=-6$ meV,  ${\cal C}=-1.4$ meV, $\mathcal{B}=-78.3$ eV\AA$^2$, $\mathcal{D}=-18$ eV\AA$^2$ and $\mathcal{A}=0.62$ eV\AA   \ \cite{Liu13_supp}.
We have used the bulk values for the $g$-factors $g_e=-14.7$ and $g_h = -9.3$.  For $B<B_{\rm cross}$ the spin-resolved Chern numbers are $C_\uparrow=-C_{\downarrow}=1$ (helical edge modes) whereas for $B>B_{\rm cross}$ the spin-resolved Chern numbers are $C_\uparrow=C_{\downarrow}=0$ (trivial insulator). 
}\label{fig:landau}
\end{figure}

We consider bilayer QSH systems, such as InAs/GaSb bilayers, described by the  BHZ Hamiltonian \cite{Bernevig06_supp, Liu08_supp}
\begin{eqnarray}
H &=&  \big\{{\cal{M}} - \mathcal{B} \big[(k_x-\frac{y}{l_B^2})^2+k_y^2\big]\big\} \sigma_0 \tau_z  + \mathcal{A} (k_x-\frac{y}{l_B^2}) \sigma_z \tau_x  - \mathcal{A} k_y \sigma_0 \tau_y + \big\{{\cal C}-\mathcal{D}\big[(k_x-\frac{y}{l_B^2})^2+k_y^2\big] \big\}\sigma_0 \tau_0 \nonumber \\ &&+\frac{g_e \mu_B B}{4} \sigma_z (\tau_0+\tau_z)+\frac{g_h \mu_B B}{4} \sigma_z (\tau_0-\tau_z), \label{eq:single_particle}
\end{eqnarray}
where $\sigma$ and $\tau$ are Pauli matrices in spin and electron-hole basis correspondingly, ${\cal{M}}$ describes the distance between the bottoms of electron and hole bands (in the inverted regime ${\cal{M}}<0$), ${\cal C}$ is the chemical potential, $\mathcal{B}, \mathcal{D}<0$ ($|\mathcal{B}|>|\mathcal{D}|$) determine the effective masses for the electron and hole bands, $g_{e(h)}$ are the $g$-factors for electron and hole bands, respectively, and the magnetic field has been written in the Landau gauge with $l_B=\sqrt{\hbar/eB}$ being the magnetic length. 
This Hamiltonian describes an electron-hole bilayer, where the electron band in one of the layers is made out of $s$-orbitals and the hole band in the other layer is made out of $p$-orbitals, so that the tunneling between the layers (proportional to $\mathcal{A}$) is odd in momentum.
There exist two strategies for constructing this kind of bilayer system. The first possibility is to use two different semiconducting materials where the $s$-like electron band in one of the materials and $p$-like hole band in the other material are inverted  such as InAs/GaSb bilayers  \cite{Liu08_supp, Du15_supp, Spanton14_supp}.
The second strategy is to use in both layers the same semiconductor where the $s$-like electron band and the $p$-like hole band are close  in energy, so that in a gated device one can reach a situation where the $s$-like electron band is active close to Fermi energy in one layer whereas the $p$-like hole band is active in the other. A promising approach to realize this  possibility is to construct a bilayer in such a way that each layer individually supports the QSH effect \cite{Michetti_supp}. Here we have neglected the spin-orbit coupling terms arising due to structural and bulk inversion asymmetry. In InAs/GaSb bilayers these terms are estimated to be very small \cite{Liu13_supp}.
The Landau level spectrum for InAs/GaSb bilayers is shown in Fig.~\ref{fig:landau}. In this material the parameters of the model can be tuned with the help of gate voltages and widths of the quantum wells. Here we have used characteristic values of the parameters \cite{Liu13_supp} and the bulk values for the $g$-factors.

The two lowest Landau level wavefunctions for this model are
\begin{align}
\psi_{k;\uparrow (\downarrow)} = \frac{e^{ikx}}{\sqrt{L l_B}}\phi\left(\frac{y - kl_B^2}{l_B}\right) \begin{pmatrix}
1(0) & 0 & 0 & 0(1)
\end{pmatrix}^T, \label{wavef_supp}
\end{align}
where $\phi(\xi) = e^{-\xi^2/2}/\pi^{1/4}$. Notice that spin and orbital degrees of freedom are locked, so that the pseudospin $\uparrow$ ($\downarrow$) means simultaneously up (down) spin and upper (lower) layer. This locking is caused by the tunneling term proportional to $\cal{A}$, which is linear in momentum. Due to the existence of this term the electron-like Landau level with spin down (hole-like Landau level with spin up)  couples to a hole-like (electron-like) higher Landau level, and as a result of this coupling these Landau levels are well separated in energy from the Landau levels  given by Eq.~(\ref{wavef_supp}).
Within the subspace generated by these wave functions, the single-particle Hamiltonian is
\begin{equation}
\hat{H}_0=\sum_{k} [\hat{\psi}_{k, \uparrow}^\dag \hat{\psi}_{k, \uparrow} - \hat{\psi}_{k, \downarrow}^\dag \hat{\psi}_{k, \downarrow} ] E_G(kl_B^2).
\label{LLham_supp}
\end{equation}
Here $\hat{\psi}_{k, \uparrow(\downarrow)}^\dag$ and $\hat{\psi}_{k, \uparrow(\downarrow)}$ are the creation and annihilation operators corresponding to the electronic states described by Eq.~(\ref{wavef_supp}), and we have fixed the chemical potential so that the total density corresponds to one of these Landau levels being filled and the other empty ($\nu_T=\nu_\uparrow+\nu_\downarrow=1$). The Fermi level is set to be at zero energy.  For an infinite system in $y$-direction, the energy $E_G$ is independent on momentum and given by $E_{Gb} = \mathcal{M} -  \frac{e \mathcal{B}}{\hbar} B+ \frac{g_e + g_h}{4}\mu_B B$.  Due to the presence of an edge the Landau levels obtain an energy-momentum dispersion. According to Eq.~(\ref{wavef_supp}) the momentum is directly connected to the position $y$ in real space, so that this energy-momentum dispersion can also be written as a position-dependent energy  $E_G(y)=E_G(k l_B^2)$. The Landau level originating from the electron (hole) band always disperses upwards (downwards) in energy, when approaching the edge. Moreover, because the edge states exist at all energies between the lowest Landau levels and the higher ones, close to the edge $E_G(y)>0$ reaches extremely large values, which are on the same order as the energy separation between the bulk Landau levels.
The spatial variation of $E_G(y)$ occurs within a characteristic length scale $l_0$, which for clean sharp edge is given by $l_0  \sim  l_B$. However, because the edge state velocity $v_s=\frac{1}{\hbar} \frac{dE_G}{dk}$ in the non-interacting theory is directly related to $l_0$, the edge roughness and disorder renormalize $l_0$ upwards ($v_s$ downwards), and hence $l_0$ can be considerably larger than $l_B$. The explicit form of the wave functions is given by Eq.~(\ref{wavef_supp}) only if $l_0 \gg l_B$, but our main results are expected to remain valid even if this condition is not satisfied.

Assuming that there is a band inversion at zero magnetic field (${\cal{M}},\mathcal{B}<0$), there is a crossing of the lowest Landau levels at magnetic field
$B_{\rm cross}= {\cal M}/(\frac{e {\cal B}}{\hbar}-\frac{g_e +g_h}{4} \mu_B)$, where the band inversion is removed.  For $B<B_{\rm cross}$, we notice that $E_G(y)=E_{Gb}<0$ in the bulk but $E_G(y)>0$ close to the edge, yielding helical edge states. On the other hand for $B>B_{\rm cross}$, $E_G(y)>0$ everywhere, and therefore the edge is gapped according to the non-interacting theory. The change in the edge structures shows up in the spin resolved Chern numbers. Namely  for $B<B_{\rm cross}$  the spin-resolved Chern numbers are $C_\uparrow=-C_{\downarrow}=1$ (helical edge modes), whereas for $B>B_{\rm cross}$ $C_\uparrow=C_{\downarrow}=0$ (trivial insulator).

\section{Energy functional}

Within the subspace generated by the lowest Landau level wave functions, the projected Hamiltonian can be written as $\hat{H}=\hat{H}_0+\hat{H}_I$, where the interactions are described by
\begin{equation}
\hat{H}_I= \frac{1}{2} \sum_{\sigma, \sigma'} \sum_{k, k', q} V_P^{\sigma \sigma'}(k-k', q) \hat{\psi}_{k\sigma}^\dag \hat{\psi}_{k', \sigma'}^\dag \hat{\psi}_{k'+q \sigma'} \hat{\psi}_{k-q \sigma}.
\end{equation}
Here, the projected Coulomb interactions can be written as 
\begin{eqnarray}
V_P^{\sigma \sigma'}(k-k' ,q)= \frac{\pi^{1/4}}{\sqrt{2}} \phi_0^2\bigg(\frac{q l_B}{\sqrt{2}} \bigg) \frac{1}{L l_B} \int d^2r  \ V^{\sigma \sigma'}(\mathbf{r}) e^{iq x}    \phi_0\bigg(\frac{y-(k-k'-q)l_B^2}{l_B} \bigg). 
\end{eqnarray}
To simplify the expressions we assume that the quantum wells are very narrow so that
\begin{equation}
V^{\uparrow\uparrow}(\mathbf{r})=V^{\downarrow\downarrow}(\mathbf{r})=\frac{e^2}{4 \pi \epsilon \epsilon_0 r}, \hspace{0.5 cm} 
V^{\uparrow\downarrow}(\mathbf{r})=V^{\downarrow\uparrow}(\mathbf{r})=\frac{e^2}{4 \pi \epsilon \epsilon_0 \sqrt{r^2+d^2}}. \label{V12}
\end{equation}
This assumption does not change the results qualitatively. However, one should keep in mind that quantitatively the energy scales associated with the interaction effects are overestimated, because the finite width of the quantum well would reduce the effective interaction strengths. 

To compute the energy for a pseudospin texture we follow closely the approach developed in Ref.~\onlinecite{Moon95_supp}. We assume that the components of the pseudospin $h_i(y)=h_i(kl_B^2)$ are slowly varying with respect to $l_B$, so that we can express the many particle wave function as 
\begin{equation}
| \Psi [\mathbf{h}(k l_B^2)]\rangle=  \prod_{k} \frac{1}{\sqrt{2[1-h_z(k l_B^2)]}} \bigg\{\big[h_x(kl_B^2)-ih_y(kl_B^2)\big]\hat{\psi}_{k, \uparrow}^\dag+\big[1-h_z(kl_B^2)\big]\hat{\psi}_{k, \downarrow}^\dag \bigg\} |0\rangle.
\end{equation}
Here $\sum_i h_i^2(k l_B^2)=1$. The energy functional can be obtained by calculating 
\begin{equation}
E\big[ \mathbf{h}(k l_B^2)\big]=\langle \Psi \big[ \mathbf{h}(k l_B^2)\big] | \hat{H}_0 + \hat{H}_I |  \Psi \big[ \mathbf{h}(k l_B^2)\big] \rangle.
\end{equation}
Using the Wick's theorem, we obtain
\begin{equation}
E\big[ \mathbf{h}(k l_B^2)\big]=E_0+\sum_k E_G(k l_B^2) h_z(kl_B^2) 
-  \sum_{k, k'}  \bigg\{ V_P^Z(k-k')  h_z(k l_B^2)h_z(k' l_B^2)  
   +  \sum_{i=x,y} V_P^{XY}(k-k') h_i(k l_B^2) h_i(k' l_B^2)  \bigg\}, \label{general-energy}
\end{equation}
where
\begin{eqnarray}
V^Z_P(q)&=&\frac{1}{4} \bigg[-V_P^{\uparrow\uparrow}(q, 0)+V_P^{\uparrow\uparrow}(q, q)+V_P^{\uparrow \downarrow}(q, 0) \bigg] 
= -\frac{V_C}{4 \sqrt{2 \pi} L}  \int_{-\infty}^{\infty}  dy \ln\bigg[\frac{y^2+d^2}{y^2} \bigg]  e^{-(y/l_B-ql_B)^2/2} \nonumber \\ &&+\frac{V_C}{4\sqrt{2 \pi} L} e^{-q^2 l_B^2/2}  \int_{-\infty}^\infty dx  \  e^{iq x}   e^{x^2/4 l_B^2} K_0\bigg(\frac{x^2}{4l_B^2}\bigg)
\end{eqnarray}
and
\begin{equation}
V_P^{XY}(q)=\frac{1}{4} V_P^{\uparrow\downarrow}(q, q)=\frac{ e^{-q^2 l_B^2/2} V_C}{4 \sqrt{2 \pi} L}    \int_{-\infty}^\infty dx  \  e^{iq x}   e^{x^2/4 l_B^2+d^2/4l_B^2} K_0\bigg(\frac{x^2+d^2}{4l_B^2}\bigg).
\end{equation}
Here the characteristic energy scale of the Coulomb interactions is $V_C=e^2/(4 \pi \epsilon \epsilon_0 l_B)$.

\section{Mean field solutions in the bulk}

Before describing the pseudospin texture at the edge, let's solve the ground state orientation of the pseudospin in the bulk. By assuming a homogeneous solution in the bulk, we obtain
\begin{eqnarray}
E[\mathbf{h}]&=&E_0+\frac{L^2}{2\pi l_B^2}   \bigg[ E_{G b} h_z  - V^Z_0 h_z^2  
   -V^{XY}_0 (h_x^2+h_y^2) \bigg], \label{meanfield}
\end{eqnarray} 
where $E_{G b}$ is the bulk value of $E_{G}(y)$ (in the inverted regime $E_{G b}<0$ and in the non-inverted regime $E_{G b}>0$),
\begin{equation}
V^Z_0 =  \sum_{q} V_P^Z(q)= V_C \frac{1}{4} \bigg[   \sqrt{\frac{\pi}{2}} - \frac{d}{l_B} \bigg]\end{equation}
and
\begin{equation}
V^{XY}_0=\sum_{q} V_P^{XY}(q)= V_C  \frac{1}{4} \sqrt{\frac{\pi}{2}}  e^{d^2/2l_B^2}  \textrm{Erfc}\bigg[\frac{d}{\sqrt{2}l_B}\bigg].
\end{equation}

By minimizing the energy given Eq.~(\ref{meanfield}) with a constraint $\sum_i h_i^2=1$, we obtain
\begin{equation}
h_{zb} = \begin{cases}
    1, &  -\frac{E_{Gb}}{2(V^{XY}_0-V^Z_0)}>1\\
     -\frac{E_{Gb}}{2(V^{XY}_0-V^Z_0)},  & \big|\frac{E_{Gb}}{2(V^{XY}_0-V^Z_0)}\big|<1 \\
     -1, &  -\frac{E_{Gb}}{2(V^{XY}_0-V^Z_0)}<-1\\
\end{cases} \label{meanfieldsol}
\end{equation}
The other components satisfy $h_{xb}^2+h_{yb}^2=1-h_{zb}^2$ and the energy is degenerate with respect to the rotations in the $(x,y)$-plane. 

To estimate the single particle excitation gap in the bulk, we can construct a mean field Hamiltonian 
by Hartree-Fock linearization of the interaction terms. This way we obtain
\begin{equation}
\hat{H}_{\rm mf}=\sum_{k} (\hat{\psi}_{k, \uparrow}^\dag, \hat{\psi}_{k, \downarrow}^\dag ) \begin{pmatrix}  E_{Gb}+m_{zb} & m_{xb}-im_{yb}  \\ m_{xb} +i m_{yb} & -E_{Gb}-m_{zb} \end{pmatrix} \begin{pmatrix} \hat{\psi}_{k, \uparrow}\\ \hat{\psi}_{k, \downarrow} \end{pmatrix}, 
\label{LLham_Supp_1}
\end{equation}
where $m_{zb}=-2 V^Z_0 h_{zb}$, $m_{xb}=-2 V^{XY}_0 h_{xb}$ and $m_{yb}=-2 V^{XY}_0 h_{yb}$. The bulk gap for single particle excitations is thus 
\begin{equation}
E_{\rm gap, s}=2 \sqrt{(E_{Gb}-2 V^Z_0 h_{zb})^2+ 4 (V^{XY}_0)^2 (h_{xb}^2+h_{yb}^2)}.
\end{equation}

In addition to the crossing of the Landau levels as a function of magnetic field, several other conditions need to be satisfied in order to realize the helical exciton condensate phase: (i) The other Landau levels at the crossing point are separated in energy so that they are not excited. (ii) The densities close to the charge neutrality point can be obtained so that $\nu_{\uparrow}$ and $\nu_{\downarrow}$ can be controlled with magnetic field and gate voltages. (iii) The layer separation described by $d/l_B$ can be made sufficiently small to reach the exciton condensate phase. (iv) The temperature can be made small enough to reach the exciton condensate phase. (v) The disorder should not be too strong.

(i) As demonstrated in Fig.~\ref{fig:landau} the typical energy gap to higher Landau levels in InAs/GaSb bilayers is on the order of $10$ meV, which is significantly larger than the gap opened by the exciton condensate order parameter. We also point out that although the large gap to higher Landau levels simplifies the theoretical analysis, it may not be necessary for the existence of exciton condensate state because the interaction effects actually tend to enhance this energy gap further. In fact, in GaAs bilayers the lowest Landau levels are separated from the higher ones by the Zeeman energy which is a rather small energy scale, but nevertheless the spin is fully polarized in the exciton condensate state \cite{Spielman-spin,Giudici,Finck10, Tiemann15}. 
 
(ii) In InAs/GaSb bilayers the densities close to the charge neutrality point can be experimentally reached in gated devices both in the absence \cite{Du15_supp, Spanton14_supp, Qu15_supp, Du15EC_supp} and in the presence \cite{Nichele14_supp} of the magnetic field. 
 
(iii) In GaAs quantum Hall bilayers it is known that the exciton condensate phase appears for $d/l_B \lesssim 1.8$
\cite{Murphy94_supp, Spielman00_supp, Kellogg02_supp, Champagne08_supp, Tiemann09_supp}. Using the typical parameters of the experimental samples 
\cite{Du15_supp, Spanton14_supp, Qu15_supp} we find that this condition can be easily satisfied in InAs/GaSb bilayers.  

(iv) The temperature needs to be smaller than the energy gap opened by the exciton condensate order parameter.  Moreover, since we are studying a two dimensional system the actual transition to the exciton condensate phase is a Berezinskii-Kosterlitz-Thouless transition \cite{Moon95_supp}, and the transition temperature can be estimated to be on the order of Kelvin. The transition temperatures measured in GaAs bilayers are consistent with this type of estimate \cite{Champagne08_supp}. 

 (v) The mean free path should be long compared to the coherence length. In quantum Hall exciton condensates the coherence length is on the order of $l_B$ and therefore this condition is easily satisfied. However, the disorder plays also another role in this system. Namely, the vortices are charged and therefore they can be nucleated by a sufficiently strong disorder potential \cite{Eastham09_supp, Sun10_supp}. Although the detailed consideration of the disorder is not the subject of this paper, we point out that most of the phenomenology of the quantum Hall exciton condensate state survives at least approximately also in the presence of the disorder-nucleated vortices \cite{Stern11_supp, Fertig05_supp, Eastham10_supp, Hyart11_supp, Hyart13_supp}. 

\section{Domain wall at the edge}

In order to describe the domain wall at the edge we first write $h_z(k l_B^2)=\cos [\theta(k l_B^2)]$, $h_{x}(k l_B^2)=\sin [\theta(k l_B^2)]$ and $h_{y}(k l_B^2)=0$. [There is a degeneracy in ($h_x, h_y$)-plane, so that we can choose a specific direction arbitrarily.]  The energy functional [Eq.~(\ref{general-energy})] can then be written as
\begin{equation}
 E\big[ \theta(k l_B^2)\big]=\sum_k E_G(k l_B^2) \cos [\theta(k l_B^2)] 
-  \sum_{k, k'}  \bigg\{ V_P^Z(k-k')  \cos [\theta(k l_B^2)] \cos [\theta(k' l_B^2)]   
   +  V_P^{XY}(k-k') \sin [\theta(k l_B^2)] \sin [\theta(k' l_B^2)]  \bigg\}.
\end{equation}
By minimizing this with respect to $\theta(k l_B^2)$, we obtain
\begin{equation}
\tan(\theta(k l_B^2))=\frac{2 \sum_{k'} V_P^{XY}(k-k') \sin [\theta(k' l_B^2)]}{2 \sum_{k'} V_P^{Z}(k-k') \cos [\theta(k' l_B^2)]-E_G(k l_B^2)}. \label{domwalleq}
\end{equation}
This can be solved numerically by iterations.

Close to the edge $E_G(y)>0$ takes large values, because the edge states are topologically protected to exist at all energies between the lowest Landau levels and higher ones. Therefore close to the edge $\theta(y)=\pi$. On the other hand, in the bulk $\theta(y)=\theta_b=\arccos(h_{zb})$. This means the existence of a domain wall, where $\theta$ rotates from $\pi$ to $\theta_b$, is a robust topological property of the system, which is not sensitive to the details of the sample.

The width of the domain wall is $l_{dw} \sim {\rm{max}}\{l_s, l_0\}$, where $l_0$ is the length scale where $E_G(y)$ changes in the vicinity of the edge
and $l_s$ is an intrinsic length scale, which is determined by the balance between the energy gain obtained by rotating $\theta$ and the corresponding loss of the exchange energy. A rough estimate is obtained by expanding Eq.~(\ref{meanfield}) around $\theta_b$ and noticing that the loss of exchange energy is approximately determined by (see Section \ref{Charged exc})
\begin{equation}
V^{XY}_2=\frac{1}{2} \sum_{q} V_P^{XY}(q) q^2 l_B^2= V_C \frac{1}{16}  \sqrt{\frac{\pi}{2}}  e^{d^2/2l_B^2}  \textrm{Erfc}\bigg[\frac{d}{\sqrt{2}l_B}\bigg]. \label{VXY2}
\end{equation}
This leads to a simple estimate 
\begin{equation}
l_s \approx \sqrt{\frac{2 V_2^{XY}}{2(V_0^{XY}-V_0^Z)(1-2h_{zb}^2)-E_{Gb} h_{zb}}} l_B.
\end{equation}
This length scale $l_s$ diverges at the phase transitions $h_z=1$, $E_{Gb}=-2(V_0^{XY}-V_0^Z)$ and $h_z=-1$, $E_{Gb}=2(V_0^{XY}-V_0^Z)$ indicating that $\theta(y)$ approaches the bulk value in a power-law like fashion (outside the phase transitions it approaches the bulk value exponentially).  Deep inside the helical quantum Hall exciton condensate phase $l_s \sim l_B$.

\section{Charged edge excitations \label{Charged exc}}

We now consider excitations  $h_z(\mathbf{r})=h_z^0(y)+\delta h_z(\mathbf{r})$, $h_x(\mathbf{r})=\sqrt{1-h_z^2(\mathbf{r})} \cos[\phi(\mathbf{r})]$ and $h_y(\mathbf{r})=\sqrt{1-h_z^2(\mathbf{r})}\sin[\phi(\mathbf{r})]$, which are described by slow spatial variation of the pseudospin within the $(x,y)$-plane [$\phi(\mathbf{r})$] and small fluctuations [$\delta h_z(\mathbf{r})$] of the $z$-component around the ground state configuration $h_z^0(y)=\cos [\theta_0(y)]$. 

For this purpose, we first write Eq.~(\ref{general-energy})
\begin{equation}
E\big[ \mathbf{h}(k l_B^2)\big]=E_0[h_z(kl_B^2)]
-  \sum_{k, k'}  \bigg\{  \sum_{i=x,y} V_P^{XY}(k-k') h_i(k l_B^2) h_i(k' l_B^2)  \bigg\}.
\end{equation}
By expanding ($i=x,y$)
\begin{equation}
h_i (k l_B^2) h_i (k' l_B^2)=h_i^2 (k l_B^2)+ h_i (k l_B^2) (k'l_B^2-k l_B^2)h'_i (k l_B^2) + \frac{1}{2} h_i (k l_B^2) (k'l_B^2-k l_B^2)^2 h''_i (k l_B^2)...  
\end{equation}
we obtain
\begin{eqnarray}
 E\big[\mathbf{h}\big] 
&=&E_0[h_z] -\frac{L}{2\pi l_B^2}\int dy \bigg\{V^{XY}_0 \sum_{i=x,y} h_i^2 (y)  + V^{XY}_2 l_B^2 \sum_{i=x,y} h_i (y)  h''_i (y) 
  \bigg\} \nonumber \\
&=& E_0[h_z] - \frac{1}{2\pi l_B^2}\int d^2 r \bigg\{   V^{XY}_0 \sum_{i=x,y} h_i^2 (\mathbf{r}) + V^{XY}_2 l_B^2 \sum_{i=x,y} h_i (\mathbf{r})  \nabla^2 h_x (\mathbf{r})  \bigg\} \nonumber \\
&=& E_0[h_z] + \frac{1}{2\pi l_B^2}\int d^2 r \bigg\{  V^{XY}_2 l_B^2 \sum_{i=x,y} \big[\nabla h_i (\mathbf{r})\big]^2  -V^{XY}_0 \sum_{i=x,y} h_i^2 (\mathbf{r})   \bigg\}.  \label{exp1}
\end{eqnarray}
 Here, we have transformed from momentum space $k$ to real space $y$ by using $y=k l_B^2$,  restored  the possibility that  $\mathbf{h}(\mathbf{r})$ may also vary in the $x$-direction (using rotational symmetry of the system), assumed a boundary condition
 \begin{equation}
 \int d^2 r \ \nabla \cdot (h_i \nabla h_i)=0
 \end{equation}
and $V^{XY}_2$ is given by Eq.~(\ref{VXY2}).

Similar gradient expansion cannot be done for $h_z(\mathbf{r})$ because of the long-range Hartree interactions. However, deep inside the phases $\delta h_z(\mathbf{r})$ mode is gapped by a large energy gap, and therefore we assume that as a first approximation it is enough to take the Hartree interactions into account only via the $V_0^Z$ term. In order to calculate the excitation energy we define a functional 
\begin{equation}
E\big[\mathbf{h}, \lambda \big]= E\big[\mathbf{h}\big] + \frac{1}{2\pi l_B^2} \int d^2r \ \lambda(\mathbf{r})(1 - \sum_{i} h_i(\mathbf{r})^2),
\end{equation}
where we have now explicitly implemented the constraint $\sum_i h_i^2=1$ with the help of Lagrange multiplier. The minimization of $E\big[\mathbf{h}, \lambda \big]$ with respect to all variables leads to a saddle point equation (\ref{domwalleq}) and allows us to consider fluctuations $\delta h_i(\mathbf{r})$ around the saddle point independently on each other. For general $E_G(y)$ the saddle point equation can be solved only numerically. Here we proceed by assuming that $E_G(y)$ is slowly varying. This is always satisfied in the bulk and we will comment the influence of the edge corrections below.   

With these approximation the excitation energy (energy difference compared to the ground state)  can be written as
\begin{equation}
\delta E\big[\delta h_z(\mathbf{r}),\phi(\mathbf{r})\big] 
= \frac{1}{2\pi l_B^2}\int d^2 r \bigg\{  V^{XY}_2 l_B^2 \sin^2[\theta_0(y)] \big[\nabla \phi(\mathbf{r})\big]^2  + \bigg[(V_0^{XY}-V_0^Z) \sin^2[\theta_0(y)]-\frac{E_G(y)}{2}\cos[\theta_0(y)] \bigg] \delta h_z^2(\mathbf{r})   \bigg\}. \label{eq:excitationenergy} 
\end{equation}

A variational wave function describing a general pseudospin texture $\mathbf{h}(\mathbf{r})$ is no longer restricted to the lowest Landau level. Therefore, the many particle wave function needs to be projected to the lowest Landau level with an operator $\cal{P}$, which can be implemented as explained in Ref.~\onlinecite{Moon95_supp}. This results in a charge density for such kind of excitations, which can be computed from the general relationship between the electric and topological charge densities \cite{Moon95_supp} 
\begin{equation}
\delta \rho(\mathbf{r})=-\frac{e}{8 \pi} \epsilon_{\mu \nu} \mathbf{h}(\mathbf{r}) \cdot [\partial_{\mu} \mathbf{h}(\mathbf{r}) \times \partial_\nu \mathbf{h}(\mathbf{r})]. \label{spin-charge-supp-matt}
\end{equation}
For the low-energy excitations, the charge density is therefore (see also Section \ref{sec:spin-charge})
\begin{equation}
\delta \rho(\mathbf{r})=-\frac{e}{4 \pi} \frac{\partial \phi}{\partial x} \frac{\partial \cos[\theta_0(y)]}{\partial y}.\label{eq:charge_densitySM}
\end{equation}
From this expression we see that the charge density is localized in the vicinity of the edge, and the length scale in $y$-direction is determined by the domain wall width $l_{dw}$. Moreover, the low-energy charged excitations are always associated with spatial variation of $\phi$ along the edge, so that the charge density is proportional to $\partial_x \phi$. We will study the nature of these excitations separately in the two opposite limits $E_{Gb} \ll - 2(V^{XY}_0-V^Z_0)$ (deep inside the uncorrelated helical quantum Hall state) and  $|E_{Gb}| \ll  2(V^{XY}_0-V^Z_0)$ (deep inside the helical quantum Hall exciton condensate phase).

\section{Charged edge excitations in a closed system}

We consider narrow quantum Hall systems with width $W$ much smaller than the length $L$ (similarly as shown in Fig.~4 in the main text), and we assume that the system is closed by connecting the ends of the sample. Based on Eqs.~(\ref{eq:excitationenergy}) and (\ref{eq:charge_densitySM}) we argue that the lowest energy charged excitations are obtained by letting azimuthal angle $\phi(x)$ rotate along the $x$-direction. In a closed system the angle $\phi(x)$ needs to rotate integer multiple of $2\pi$. 
The lowest energy excitations in a closed system, which carry a net charge within one of the edges correspond to rotation of $\phi(x)$ by $2\pi$, and they have an energy 
$\delta E \sim V_2^{XY} l_{dw}/L$ deep inside the uncorrelated phase and $\delta E \sim V_2^{XY} W/L$ deep inside the helical exciton condensate phase. 

In the uncorrelated phase we obtain from Eq.~(\ref{eq:charge_densitySM}) that these excitations have a charge $\pm e$. Moreover, they are deconfined: One can create these excitations independently on the different edges.

In the helical quantum Hall exciton condensate phase, we obtain from Eq.~(\ref{eq:charge_densitySM}), that the elementary excitations in a closed system have a charge is $\pm \nu_\uparrow e$, where $\nu_{\uparrow}=(1+\cos\theta_b)/2$. Moreover, a charge  $\pm \nu_{\uparrow}e$ on one of the edges is always connected to the opposite charge on the other edge by a stripe of rotated pseuspins through the bulk. Creating this stripe costs an energy  $\delta E \sim V_2^{XY} W/L$, which is proportional to the distance $W$ between the charges. However, breaking the stripe somewhere in the bulk would cost a much larger energy comparable to the Coulomb energy. Thus isolated charges cannot be observed at low energies. This means that in the helical quantum Hall exciton condensate phase the charged edge excitations are confined.

\section{Luttinger liquid theory}

To develop time-dependent theory for these excitations, we use the known result that the Euclidian action in the adiabatic approximation can be written as \cite{Moon95_supp}
\begin{equation}
S^E[\mathbf{h}, \lambda]= \int_0^\beta d\tau \bigg\{ \int d^2r \bigg[-i \frac{1}{4 \pi l_B^2} \mathbf{A}(\mathbf{h})\cdot \partial_\tau \mathbf{h} \bigg]+E\big[\mathbf{h}, \lambda \big]  \bigg\},
\end{equation}
where $\nabla_\mathbf{h} \times \mathbf{A}(\mathbf{h})=\mathbf{h}$.
Thus by assuming that $E_G(y)$ is slowly varying function, we can expand the action around the saddle point and obtain 
\begin{align}
\delta S^E[\phi, \delta h_z] = \frac{1}{2\pi l_B^2} \int_0^\beta d\tau \int d^2r \bigg\{&
\frac{i}{2} \delta h_z \frac{\partial \phi}{\partial \tau} + V_2^{XY} l_B^2 \sin^2 \theta_0 +  \bigg[(V_0^{XY}-V_0^Z) \sin^2 \theta_0(y)-\frac{E_G(y)}{2}\cos \theta_0(y) \bigg]  \delta h_z^2
\bigg\}.
\end{align}

\subsection{Confined phase}

Deep inside the confined phase, all the properties of the system will be determined by the bulk (see below). Therefore, we neglect the effects arising in the vicinity of the edge, and use Eq.~(\ref{meanfieldsol}) to rewrite the action as
\begin{align}
\delta S^E[\phi, \delta h_z] = \frac{1}{2\pi l_B^2} \int_0^\beta d\tau \int d^2r \bigg\{&
\frac{i}{2} \delta h_z \frac{\partial \phi}{\partial \tau} + V_2^{XY} l_B^2 \sin^2 \theta_0 + (V_0^{XY} - V_0^Z) \delta h_z^2
\bigg\}
\end{align}

We integrate out the massive fluctuations $\delta h_z$. This way we obtain
\begin{equation}
S^E\big[\phi(\mathbf{r}, \tau)\big] = \frac{1}{2\pi l_B^2}\int_0^\beta d\tau \int d^2 r \bigg\{ \frac{ (\partial_\tau \phi)^2}{16(V_0^{XY}-V_0^Z)}+ V^{XY}_2 l_B^2 \sin^2[\theta_0(y)] \big[\nabla \phi(\mathbf{r})\big]^2     \bigg\}.  
\end{equation}

By going to real time, we identify the Langrangian density as
\begin{equation}
\mathcal{L}= \frac{\hbar^2 \Gamma}{2 e^2} \dot{\phi}^2-\frac{\rho_s(y)}{2} (\nabla \phi)^2. \label{Lagrangian}
\end{equation}
Here $\rho_s(y)=V^{XY}_2 \sin^2[\theta_0(y)]/\pi $ is the pseudospin stiffness and $\Gamma=e^2/(16 \pi l_B^2 (V_0^{XY}-V_0^Z))$ describes the interlayer capacitance per unit area, which is strongly enhanced from the electrostatic value by the exchange interactions. (As an important check we notice that if one neglects the exchange contributions in $V_0^{XY}-V_0^Z$, the expression for $\Gamma$ simplifies to the familiar electrostatic formula for the interlayer capacitance.)

The corresponding equation of motion is
\begin{align}
\frac{\hbar^2 \Gamma}{e^2} \ddot{\phi} - \nabla\left[\rho_s(y) \nabla \phi \right] = 0. \label{waveeqsupp}
\end{align}
This equation is translationary invariant in $x$ and $t$ thus we can express the solution as
\begin{align}
\phi = \phi_0 + \sum_{n, k} e^{- i [\omega_n(k) t - k x]} f_{n, k} (y), \label{trialwave}
\end{align}
where $\phi_0$ is an arbitrary constant and $f_{n, k} (y)$ describe the different transverse modes with energy-momentum dispersions $E_n(k)=\hbar \omega_n(k)$. By substituting the trial solution (\ref{trialwave}) to Eq.~(\ref{waveeqsupp}), we find that the eigenfunctions $f_{n, k} (y)$ and eigenfrequencies $\omega_n(k)$ satisfy an eigenvalue equation 
\begin{align}
-\frac{\hbar^2 \Gamma \omega_n(k)^2}{e^2} f_{n,k} (y) + k^2 \rho_s(y) f_{n,k}(y) - \rho_s(y) f_{n,k}''(y) - \rho_s'(y) f_{n,k}'(y)=0,
\end{align}
which can be solved numerically.

Numerical results for the helical quantum Hall exciton condensate phase show that the lowest energy mode $f_{0, k}(y)$ can be approximated as a constant throughout the sample. Moreover, the lowest energy mode at reasonably small momentum is separated from higher modes by an energy $\Delta E \sim \hbar v \pi/W$. We restrict the analysis to sufficiently low-energies that these higher modes are not excited. Then we assume that $\phi$ is constant in $y$-direction and the dispersion relation is
\begin{align}
\omega(k) =  \sqrt{\frac{e^2 \rho_{sb}}{\hbar^2 \Gamma}} k =  v k,
\end{align}
where $\rho_{sb}$ is the bulk value of $\rho_s(y)$. The Hamiltonian of the system then becomes
\begin{equation}
H  = \int dx \bigg[ \frac{e^2}{2 W \hbar^2 \Gamma} \Pi(x)^2 + \frac{W \rho_{sb}}{2} (\partial_x \phi(x))^2 \bigg],
\end{equation}
where $\Pi(x)$ is the momentum conjugate to $\phi(x)$. The one-dimensional charge densities in the different edges (labelled 1 and 2) are
\begin{align}
\rho_{1, 2}(x) = \mp \frac{e}{4\pi}(1+ \cos \theta_b)\frac{\partial \phi}{\partial x},
\end{align}
where $\theta_b$ is the bulk value of $\theta(y)$. The confinement of the edge excitation clearly shows up here as the the charge densities on the different edges are always opposite and determined by a single field $\phi(x)$. This Hamiltonian describes a Luttinger liquid. In the standard convention \cite{Giamarchi_supp} the Luttinger liquid theory is written as 
\begin{equation}
H  = \frac{\hbar}{2\pi}\int dx \bigg[ \frac{vK \pi^2}{\hbar^2} \tilde{\Pi}(x)^2 + \frac{v}{K} (\partial_x \tilde{\phi})^2 \bigg], \hspace{0.5cm} \rho(x) = - \frac{e}{\pi} \frac{\partial \tilde{\phi}}{\partial x}.
\end{equation}
To rewrite our Hamiltonian in this convention, we apply the transformation preserving canonical commutation relations 
\begin{equation}
\phi(x)=\frac{4}{1+\cos \theta_b} \tilde{\phi}(x), \hspace{0.5 cm} \Pi(x)=\frac{1+\cos \theta_b}{4}\tilde{\Pi}(x),
\end{equation}
and notice that the charge density at each edge is $\rho_{1, 2}(x) = \pm \rho(x)$. 

This way, we can identify the Luttinger liquid parameters $K$ and $v$ as
\begin{equation}
v=\sqrt{\frac{e^2 \rho_{sb}}{\hbar^2 \Gamma}}, \hspace{0.5 cm} K=\frac{e}{\pi W \sqrt{\Gamma \rho_{sb}}} \bigg(\frac{1+\cos\theta_b}{4}\bigg)^2=\frac{l_B}{W} \sqrt{\frac{V_0^{XY} - V_0^Z}{V_2^{XY}}} \frac{(1+\cos\theta_b)^2}{4 \sin\theta_b}.
\end{equation}
Here, $v$ is just the pseudospin wave velocity obtained earlier from the dispersion $\omega(k)$. The Luttinger parameter $K$ determines the conductance $G_{cf}=Ke^2/h$. However, it is important to notice that $G_{cf}$ describes the conductance for a counterflow/drag geometry,  where opposite currents are flowing in the two edges. The helical quantum Hall exciton condensate phase does not support net transport current as long as the voltages $eV$ are small compared to $\hbar v \pi/W$.

\subsection{Deconfined phase}

In the uncorrelated helical quantum Hall phase the transverse modes are localized within the length scale $l_{dw}$ from the edges. Therefore, in contrast to helical quantum Hall exciton condensate phase we can define independent fields $\phi_{1(2)}(x)$ and charge densities $\rho_{1(2)}(x) = \mp \frac{e}{2\pi}\frac{\partial \phi_{1(2)}}{\partial x}$. Thus charges can be created independently on the different edges, highlighting that in the uncorrelated helical  phase the charged edge excitations are deconfined. 

We can now repeat the calculation done above for the confined phase. This way, assuming that $E_G(y)$ is slowly varying and that the transverse modes are described by a constant within a distance $l_{dw}$ in the vicinity of the edge and $0$ elsewhere, we arrive to Luttinger liquid Hamiltonian  
\begin{equation}
H  = \sum_{i=1}^2 \int dx \bigg[ \frac{e^2}{2 l_{dw} \hbar^2 \Gamma_{av}} \Pi_i(x)^2 + \frac{l_{dw} \rho_{s, av}}{2} (\partial_x \phi_i(x))^2 \bigg], \label{suppLLdeconf}
\end{equation}
where
$\rho_{s, av}=\frac{1}{l_{dw}} \int_0^{l_{dw}} dy \  V^{XY}_2 \sin^2[\theta_0(y)]/\pi$  and $\Gamma_{av}=\frac{1}{l_{dw}} \int_0^{l_{dw}} dy \ e^2/\{16 \pi l_B^2 [(V_0^{XY}-V_0^Z) \sin^2 \theta_0(y)-\frac{E_G(y)}{2}\cos \theta_0(y)]\}$. 
This way the Luttinger liquid parameters are identified as
\begin{equation}
v=\sqrt{\frac{e^2 \rho_{s,av}}{\hbar^2 \Gamma_{av}}}, \hspace{0.5 cm} K=\frac{e}{4 \pi l_{dw} \sqrt{\Gamma_{av} \rho_{s,av}}}. 
\end{equation}
However, in contrast to the helical quantum Hall exciton condensate phase the Luttinger liquid parameters depend strongly on the detailed shape of the $E_{G}(y)$ function, and if it is not slowly varying in the vicinity of the edge, there can be large corrections to the expressions above. Nevertheless, the structure of the Luttinger liquid theory (\ref{suppLLdeconf}) is very general, and for physically reasonable parameters $K \sim 1$.  Thus we expect this phase to support helical edge state transport with conductance $G \sim e^2/h$. Here $G$ describes the conductance for a transport geometry, where a net transport current is flowing along one of the edges. 

\section{Spin texture-charge density relation in quantum Hall systems and numerical calculation of the electric charge density for the spin textures \label{sec:spin-charge}}

It is known that both topological and non-topological contributions to the electric charge exist at the vortices in various systems  \cite{Khomskii, Blatter, Kumagai, Natsik, Shevc, Volovik, Rukin, Adamenko}. It is possible to show that the  mechanisms considered in these systems do not give rise to a non-topological contribution to the electric charge in quantum Hall exciton condensates. However, instead of going through the mechanisms one-by-one,  we present general arguments for the topological spin texture-charge density relation and discuss the conditions for its breakdown. We complement the analytical arguments with a numerical calculation of the electric charge density for the pseudospin textures considered in the paper.

Originally the spin texture-charge density relation for quantum Hall systems was argued from the Chern-Simons relation between the density and the statistical magnetic field \cite{Sondhi93_supp}. Namely, the electrons feel the spin texture via the additional Berry's phase and therefore the orbital degrees of freedom are influenced in the same way as if additional magnetic flux density was inserted into the system. The extra magnetic flux is associated with an extra charge yeilding to Eq.~(\ref{spin-charge-supp-matt})  \cite{Sondhi93_supp, Girvin-review_supp}. This argument is valid only if the spin textures are smooth  on the scale of $l_B$. There exist also an alternative argument for the special case where the filling factors are $\nu_{\uparrow}=\nu_{\downarrow}=1/2$ and one neglects the higher Landau levels, so that there is a particle-hole symmetry in the system. In that case there is a general argument that a localized charge bound to a defect must be a multiple of $\pm e/2$ \cite{Hou07_supp}. Therefore, the possible charges $\pm e/2$ appearing in the special case $\nu_{\uparrow}=\nu_{\downarrow}=1/2$ can be explained this way \cite{Girvin-MacDonald-review_supp}. If the higher Landau levels are excited the particle hole symmetry is no longer exact. Therefore, this argument again relies on smoothness of the spin textures. Finally, the Eq.~(\ref{spin-charge-supp-matt}) can be obtained with an explicit microscopic calculation \cite{Moon95_supp} and it can be demonstrated for specific spin textures with the help of explicit construction of the many-particle wave functions \cite{Girvin-MacDonald-review_supp}. These approaches also rely on the assumption that the spin textures are smooth on the scale of $l_B$.

Based on these arguments  it is clear that the relation between the pseudospin texture and the charge density [Eq.~(\ref{spin-charge-supp-matt})] is valid if the pseudospin textures are smooth on the scale of $l_B$. The pseudospin textures considered in the main text  [see Fig.~4 in the main text] satisfy this requirement. 
In fact we can explicitly construct the many-particle wave functions to illustrate how the charges $\pm \nu_\uparrow e$ appear in these textures. Namely, consider a finite system with length $L$  and width $W$ illustrated in Fig.~4 in the main text. Because the momentum $k$ in the Landau level wave functions is directly connected to the position $y$ in the real space $y=k l_B^2$, the possible values of $k$ are $k_n=n 2\pi/L$, where $n=0,1,...,N$ and $N=WL/(2\pi l_B^2)$. The ground state pseudospin texture is given by  $h_z(y)=h_z(k l_B^2)=\cos [\theta_0(k l_B^2)]$, $h_x(y)=h_{x}(k l_B^2)=\sin [\theta_0(k l_B^2)]$, $h_y(y)=0$ so that the many particle wave-function for the ground state $| \Psi \rangle_{\rm GS}$ can be written as
\begin{equation}
| \Psi \rangle_{\rm GS}=  \prod_{n=0}^N \frac{1}{\sqrt{2\{1-\cos [\theta_0(k_n l_B^2)]\}}} \bigg\{\sin [\theta_0(k_n l_B^2)] \hat{\psi}_{k_n, \uparrow}^\dag+\big[1-\cos [\theta_0(k_n l_B^2)]\big]\hat{\psi}_{k_n, \downarrow}^\dag \bigg\} |0\rangle.
\end{equation}
The elementary excitation is described by a pseudospin texture $h_z(y)=h_z(k l_B^2)=\cos [\theta_0(k l_B^2)]$, $h_x(x,y)=h_{x}(x, k l_B^2)=\sin [\theta_0(k l_B^2)] \cos[\phi(x)]$, $h_y(x,y)=h_{y}(x, k l_B^2)=\sin [\theta_0(k l_B^2)] \sin[\phi(x)]$, where $\phi(x)=2 \pi x/L$. The winding of $\phi(x)$ can be removed by introducing a momentum shift $\Delta k=2\pi/L$ between the electron and hole Landau level wave functions. Because the pseudospin-texture is slowly varying and the pseudopin close to the edge points down the many-particle wave function for the excited state can be written as
\begin{equation}
| \Psi \rangle_{\rm ES}= \hat{\psi}_{k_{0}, \downarrow}^\dag \prod_{n=0}^{N-1} \frac{1}{\sqrt{2\{1-\cos [\theta_0(k_n l_B^2)]\}}} \bigg\{\sin [\theta_0(k_n l_B^2)] \hat{\psi}_{k_n, \uparrow}^\dag+\big[1-\cos [\theta_0(k_{n} l_B^2)]\big]\hat{\psi}_{k_{n+1}, \downarrow}^\dag \bigg\} |0\rangle. \label{MBES}
\end{equation}
From these expression one finds that the charge density is
\begin{equation}
\delta \rho(y)=-\frac{e}{4 \pi l_B^2}\bigg\{\cos\big[\theta_0(y+\frac{2\pi}{L} l_B^2)\big]-\cos\big[\theta_0(y)\big]\bigg\}=-\frac{e}{4 \pi} \frac{\partial \phi}{\partial x}\frac{\partial \cos\big[\theta_0(y)\big]}{\partial y}
\end{equation}
in agreerement with Eq.~(\ref{eq:charge_densitySM}). From this expression it straightforwardly follows that the charges appearing at each edge for the pseudospin texture shown in Fig.~4 in the main text are $\pm \nu_{\uparrow} e$, where $\nu_{\uparrow}$ is the bulk filling factor for pseudospin up. It is also clear that in a closed system the charges of the possible edge excitations and the charges of the vortices must be related. As shown in Fig.~5 in the main text it is possible to end the charged edge excitation into a bulk vortex, and the total charge in the system must be an integer multiple of $e$. This requirement is satisfied because the possible charges of the vortices are $\pm \nu_{\uparrow (\downarrow)} e$ and $\nu_{\uparrow}+\nu_{\downarrow}=1$.

In all arguments so far the whole pseudospin texture was considered to be smooth on the scale of $l_B$. However, it turns out that it is possible to relax this assumption. Namely, if one considers the type of pseudospin textures shown in Fig.~4 in the main text it is actually enough that {\it the pseudospin texture is smooth on the scale of $l_B$ in the bulk region between the localized charges}. Close to the edge it may vary arbitrarily sharply. To understand this consider first the type of smooth pseudospin texture shown in Fig.~4, such that charge $\nu_{\uparrow} e$ appears at one edge and $-\nu_{\uparrow} e$ at the other. We can now deform one end of the pseudospin texture in such a way that the rest of the pseudospin texture remains smooth. Then, there is still charge $\nu_{\uparrow} e$ localized on one edge and the bulk is charge neutral, which means that a charge $-\nu_{\uparrow} e$ necessarily remains at the other edge although it is no longer smooth on the scale of $l_B$. 
 The values of the polarization charges in this system are therefore fully topological. No local perturbation  in the pseudospin texture can change the charges appearing at the edges. Only a global perturbation which  connects the  fractionally charged excitations may give rise to redistribution of the charges and modify their values from $\pm \nu_{\uparrow} e$. 
 
 \begin{figure}
\includegraphics[width = 0.8\linewidth]{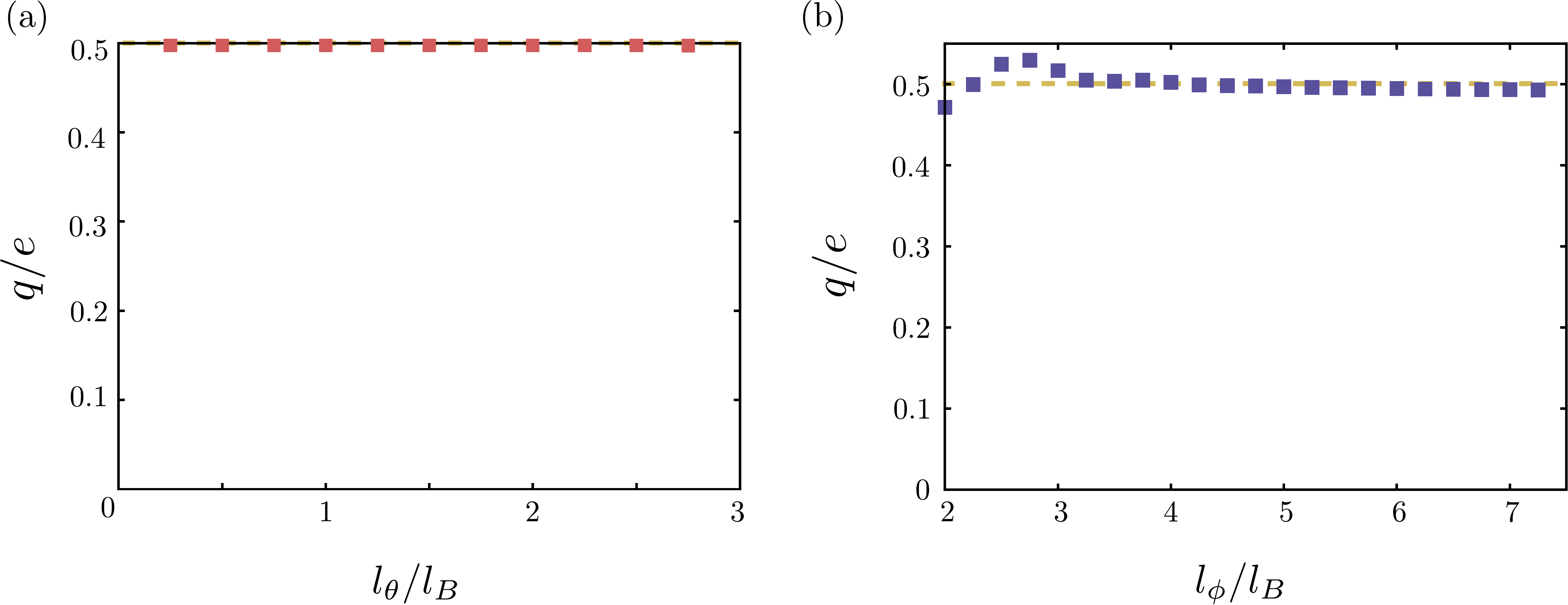}
\caption{Charge of the edge excitation in a closed system. Charge concentrated close to the edge of the sample as a function of (a) $l_\theta$ and (b) $l_\phi$. Here  $l_\theta$ determines the length scale where the angle $\theta(y)$ is varied from $\pi$ to $\theta_b$ in the vicinity of the edge and $l_\phi$ describes length scale where the angle $\phi(x)$ rotates by $2\pi$. The charge $q$ is computed by integrating the charge density over a single period $l_\phi$ along the $x$-direction and over the interval of width $5 l_B$ from the edge of the sample in $y$-direction. We have chosen $\theta_b=\pi/2$ so that the expected charge shown by a dotted line is $q=\nu_{\uparrow} e=e/2$. As expected based on the analytical arguments the value of $q$ is independent on $l_\theta$ (as long as the width of the integration interval in $y$-direction is chosen to be larger than the width of the  domain wall where the charge is concentrated in the vicinity of the edge), and deviations from the value $e/2$ as a function of $l_\phi$ start to occur only when $l_\phi \lesssim 4 l_B$.  In the numerics we have chosen the order parameter strength $\Delta_0= 1$ meV, the width of the sample $W = 20 l_B$, and the space is discretized using a lattice constant $0.25 l_B$. For (a) we fix $l_\phi=5l_B$ and for (b) $l_\theta=3 l_B$. The parameters of the BHZ Hamiltonian are same as in Fig.~\ref{fig:landau}. 
}\label{fig:charge}
\end{figure}

We have numerically verified the statements based on the analytical arguments given above. For this purpose we have inserted an order parameter term
\begin{equation}
H_{\Delta}=-\Delta_0 \bigg[h_z(x,y)\sigma_z \tau_0+h_x(x,y) \sigma_x \tau_x + h_y(x,y) \sigma_x \tau_y \bigg]
\label{orderparameter-num}
\end{equation}
to the BHZ Hamiltonian [Eq.~(\ref{eq:single_particle})]. The charge density associated with the excitation $\delta \rho(\mathbf{r})$  can be numerically computed by calcuting the difference between the charge densities for pseudospin textures  $h_z(x,y)=\cos [\theta(y)]$, $h_x(x,y)=\sin [\theta(y)] \cos[\phi(x)]$, $h_y(x,y)=\sin [\theta(y)] \sin[\phi(x)]$, where $\phi(x)=2 \pi x/l_\phi$ (excited state) and $\phi(x)=0$ (ground state). Here we have introduced a length scale $l_\phi$ where the phase $\phi(x)$ rotates over $2\pi$. The charge of an elementary excitation $q$ is obtained by integrating the charge density over a period $l_\phi$ in $x$-direction and over the width of the domain wall in $y$-direction. If we choose $l_\phi \gg l_B$ the charge of the excitation according to our analytic arguments should be $q=\pm \nu_\uparrow e$. However, $l_\phi$ allows us also to control how fast the pseudospin direction changes in such a way that it influences the system everywhere in the bulk. Therefore, if we choose $l_\phi \lesssim l_B$ it is possible to excite the higher Landau levels everywhere in the bulk so that topological protection is destroyed, and we can study how this affects the charge of the excitation. Additionally we can illustrate the topological nature of the charge of the pseudospin texture by demonstrating that deformations of the pseudospin texture appearing only close to the edge do not affect the value of $q$.  For this purpose we define 
\begin{equation}
\theta(y) = \begin{cases}
    \pi+(\theta_b-\pi)y/l_\theta, &  0<y<l_\theta \\
    \theta_b,  & l_\theta<y<W-l_\theta \\
      \theta_b+(\pi-\theta_b)(y-W+l_\theta)/l_\theta, &  W-l_\theta<y<W\\
\end{cases} \label{meanfieldsol_1}
\end{equation}
in such a way that $l_\theta$ allows to control the length scale where the angle $\theta(y)$ is varied from $\pi$ to $\theta_b$ in the vicinity of the edge. Based on our topological arguments the  charge $q$ should be independent of $l_\theta$.  (We assume $W \gg l_\theta$.)

The numerical results are shown in Figs.~\ref{fig:charge}. As can be seen from these figures for smooth pseudospin textures the numerical results are in agreement with the analytical expectations. Moreover, $q$ does not depend on $l_\theta$ [Fig.~\ref{fig:charge}(a)] in agreement with the topological argument. On the other hand,  if we deform  the pseudospin texture in such a way that it rotates fast in the bulk by choosing $l_\phi \lesssim 4 l_B$ it is possible to excite the higher Landau levels everywhere in the bulk so that topological protection is destroyed. This results in redistribution of the charges giving rise to deviations of $q$ from the value $\pm \nu_{\uparrow} e$ [Fig.~\ref{fig:charge}(b)].


\begin{thebibliography}{99}
\bibitem{comment} It is known that for sufficiently large interlayer separation the quantum Hall bilayers display an instability towards formation of a charge density wave ground state \cite{CoteBreyMacDonald92}. Here we assume that the interlayer separation is small enough that such kind of instability does not occur.
\bibitem{comment2}
  If $E_{Gb}/V_0^{XY}$ is varied with magnetic field the ratio $V_0^{XY}/V_0^Z$ will change within the phase-diagram. Nevertheless, we find that the phase-diagram stays qualitatively similar also in this case.
\bibitem{comment3} Alternatively, instead of comparing the energies to create elementary excitations in a closed system one could compare the energies needed to create a fixed charge density on the edge. This generalization allows the possibility to consider open systems.
\end{thebibliography}
\end{document}